\documentclass[twocolumn,twocolappendix]{aastex631}

\newcommand{\p}[1]{ ^{#1} }
\newcommand{\eq}[1]{\begin{equation} #1 \end{equation}} 

\newcommand{\parens}[1]{\left ( #1 \right )} 

\newcommand{\eqref}[1]{Equation~\ref{#1}}
\newcommand{\secref}[1]{Section~\ref{#1}}
\newcommand{\figref}[1]{Figure~\ref{#1}}
\newcommand{\tabref}[1]{Table~\ref{#1}}

\newcommand{\unitVec}[1]{\mathbf{\hat{#1}}} 

\shorttitle{BICEP / Keck XVI: Characterizing Dust Polarization Through Correlations with H~\textsc{i}}

\graphicspath{{./}{}}

\begin{document}

\title{BICEP / Keck XVI: Characterizing Dust Polarization through Correlations with Neutral Hydrogen}

\correspondingauthor{George Halal}
\email{georgech@stanford.edu}

\author{BICEP/Keck Collaboration: P.~A.~R.~Ade}
\affiliation{School of Physics and Astronomy, Cardiff University, Cardiff, CF24 3AA, United Kingdom}
\author[0000-0002-9957-448X]{Z.~Ahmed}
\affiliation{Kavli Institute for Particle Astrophysics and Cosmology, SLAC National Accelerator Laboratory, 2575 Sand Hill Rd, Menlo Park, CA 94025, USA}
\author{M.~Amiri}
\affiliation{Department of Physics and Astronomy, University of British Columbia, Vancouver, British Columbia, V6T 1Z1, Canada}
\author[0000-0002-8971-1954]{D.~Barkats}
\affiliation{Center for Astrophysics, Harvard \& Smithsonian, Cambridge, MA 02138, USA}
\author[0000-0002-3351-3078]{R.~Basu~Thakur}
\affiliation{Department of Physics, California Institute of Technology, Pasadena, CA 91125, USA}
\author[0000-0001-9185-6514]{C.~A.~Bischoff}
\affiliation{Department of Physics, University of Cincinnati, Cincinnati, OH 45221, USA}
\author[0000-0003-0848-2756]{D.~Beck}
\affiliation{Department of Physics, Stanford University, Stanford, CA 94305, USA}
\affiliation{Kavli Institute for Particle Astrophysics and Cosmology, SLAC National Accelerator Laboratory, 2575 Sand Hill Rd, Menlo Park, CA 94025, USA}
\author{J.~J.~Bock}
\affiliation{Department of Physics, California Institute of Technology, Pasadena, CA 91125, USA}
\affiliation{Jet Propulsion Laboratory, Pasadena, CA 91109, USA}
\author{H.~Boenish}
\affiliation{Center for Astrophysics, Harvard \& Smithsonian, Cambridge, MA 02138, USA}
\author{E.~Bullock}
\affiliation{Minnesota Institute for Astrophysics, University of Minnesota, Minneapolis, MN 55455, USA}
\author{V.~Buza}
\affiliation{Kavli Institute for Cosmological Physics, University of Chicago, Chicago, IL 60637, USA}
\author[0000-0002-1630-7854]{J.~R.~Cheshire~IV}
\affiliation{Minnesota Institute for Astrophysics, University of Minnesota, Minneapolis, MN 55455, USA}

\author[0000-0002-7633-3376]{S.~E.~Clark}
\affiliation{Department of Physics, Stanford University, Stanford, CA 94305, USA}
\affiliation{Kavli Institute for Particle Astrophysics and Cosmology, SLAC National Accelerator Laboratory, 2575 Sand Hill Rd, Menlo Park, CA 94025, USA}

\author{J.~Connors}
\affiliation{Center for Astrophysics, Harvard \& Smithsonian, Cambridge, MA 02138, USA}
\author[0000-0002-2088-7345]{J.~Cornelison}
\affiliation{Center for Astrophysics, Harvard \& Smithsonian, Cambridge, MA 02138, USA}
\author{M.~Crumrine}
\affiliation{School of Physics and Astronomy, University of Minnesota, Minneapolis, MN 55455, USA}
\author[0000-0002-7471-719X]{A.~Cukierman}
\affiliation{Department of Physics, Stanford University, Stanford, CA 94305, USA}
\affiliation{Kavli Institute for Particle Astrophysics and Cosmology, SLAC National Accelerator Laboratory, 2575 Sand Hill Rd, Menlo Park, CA 94025, USA}
\affiliation{Department of Physics, California Institute of Technology, Pasadena, CA 91125, USA}
\author{E.~V.~Denison}
\affiliation{National Institute of Standards and Technology, Boulder, CO 80305, USA}
\author{M.~Dierickx}
\affiliation{Center for Astrophysics, Harvard \& Smithsonian, Cambridge, MA 02138, USA}
\author{L.~Duband}
\affiliation{Service des Basses Temp\'{e}ratures, Commissariat \`{a} l'Energie Atomique, F-38054 Grenoble, France}
\author{M.~Eiben}
\affiliation{Center for Astrophysics, Harvard \& Smithsonian, Cambridge, MA 02138, USA}
\author[0000-0002-3790-7314]{S.~Fatigoni}
\affiliation{Department of Physics and Astronomy, University of British Columbia, Vancouver, British Columbia, V6T 1Z1, Canada}
\author[0000-0001-8217-6832]{J.~P.~Filippini}
\affiliation{Department of Physics, University of Illinois at Urbana-Champaign, Urbana, IL 61801, USA}
\affiliation{Department of Astronomy, University of Illinois at Urbana-Champaign, Urbana, IL 61801, USA}
\author{S.~Fliescher}
\affiliation{School of Physics and Astronomy, University of Minnesota, Minneapolis, MN 55455, USA}
\author{C.~Giannakopoulos}
\affiliation{Department of Physics, University of Cincinnati, Cincinnati, OH 45221, USA}
\author{N.~Goeckner-Wald}
\affiliation{Department of Physics, Stanford University, Stanford, CA 94305, USA}
\author[0000-0001-5268-8423]{D.~C.~Goldfinger}
\affiliation{Center for Astrophysics, Harvard \& Smithsonian, Cambridge, MA 02138, USA}
\author{J.~Grayson}
\affiliation{Department of Physics, Stanford University, Stanford, CA 94305, USA}
\author[0000-0001-9292-6297]{P.~Grimes}
\affiliation{Center for Astrophysics, Harvard \& Smithsonian, Cambridge, MA 02138, USA}
\author{G.~Hall}
\affiliation{School of Physics and Astronomy, University of Minnesota, Minneapolis, MN 55455, USA}
\author[0000-0003-2221-3018]{G.~Halal}
\affiliation{Department of Physics, Stanford University, Stanford, CA 94305, USA}
\author{M.~Halpern}
\affiliation{Department of Physics and Astronomy, University of British Columbia, Vancouver, British Columbia, V6T 1Z1, Canada}
\author{E.~Hand}
\affiliation{Department of Physics, University of Cincinnati, Cincinnati, OH 45221, USA}
\author{S.~Harrison}
\affiliation{Center for Astrophysics, Harvard \& Smithsonian, Cambridge, MA 02138, USA}
\author{S.~Henderson}
\affiliation{Kavli Institute for Particle Astrophysics and Cosmology, SLAC National Accelerator Laboratory, 2575 Sand Hill Rd, Menlo Park, CA 94025, USA}
\author[0000-0003-0220-0009]{S.~R.~Hildebrandt}
\affiliation{Department of Physics, California Institute of Technology, Pasadena, CA 91125, USA}
\affiliation{Jet Propulsion Laboratory, Pasadena, CA 91109, USA}
\author{J.~Hubmayr}
\affiliation{National Institute of Standards and Technology, Boulder, CO 80305, USA}
\author[0000-0001-5812-1903]{H.~Hui}
\affiliation{Department of Physics, California Institute of Technology, Pasadena, CA 91125, USA}
\author{K.~D.~Irwin}
\affiliation{Department of Physics, Stanford University, Stanford, CA 94305, USA}
\affiliation{Kavli Institute for Particle Astrophysics and Cosmology, SLAC National Accelerator Laboratory, 2575 Sand Hill Rd, Menlo Park, CA 94025, USA}
\affiliation{National Institute of Standards and Technology, Boulder, CO 80305, USA}
\author[0000-0002-3470-2954]{J.~Kang}
\affiliation{Department of Physics, Stanford University, Stanford, CA 94305, USA}
\affiliation{Department of Physics, California Institute of Technology, Pasadena, CA 91125, USA}
\author[0000-0002-5215-6993]{K.~S.~Karkare}
\affiliation{Center for Astrophysics, Harvard \& Smithsonian, Cambridge, MA 02138, USA}
\affiliation{Kavli Institute for Cosmological Physics, University of Chicago, Chicago, IL 60637, USA}
\author{E.~Karpel}
\affiliation{Department of Physics, Stanford University, Stanford, CA 94305, USA}
\author{S.~Kefeli}
\affiliation{Department of Physics, California Institute of Technology, Pasadena, CA 91125, USA}
\author{S.~A.~Kernasovskiy}
\affiliation{Department of Physics, Stanford University, Stanford, CA 94305, USA}
\author{J.~M.~Kovac}
\affiliation{Center for Astrophysics, Harvard \& Smithsonian, Cambridge, MA 02138, USA}
\affiliation{Department of Physics, Harvard University, Cambridge, MA 02138, USA}
\author{C.~L.~Kuo}
\affiliation{Department of Physics, Stanford University, Stanford, CA 94305, USA}
\affiliation{Kavli Institute for Particle Astrophysics and Cosmology, SLAC National Accelerator Laboratory, 2575 Sand Hill Rd, Menlo Park, CA 94025, USA}
\author[0000-0002-6445-2407]{K.~Lau}
\affiliation{School of Physics and Astronomy, University of Minnesota, Minneapolis, MN 55455, USA}
\author{E.~M.~Leitch}
\affiliation{Kavli Institute for Cosmological Physics, University of Chicago, Chicago, IL 60637, USA}
\author{A.~Lennox}
\affiliation{Department of Physics, University of Illinois at Urbana-Champaign, Urbana, IL 61801, USA}
\author{K.~G.~Megerian}
\affiliation{Jet Propulsion Laboratory, Pasadena, CA 91109, USA}
\author{L.~Minutolo}
\affiliation{Department of Physics, California Institute of Technology, Pasadena, CA 91125, USA}
\author[0000-0002-4242-3015]{L.~Moncelsi}
\affiliation{Department of Physics, California Institute of Technology, Pasadena, CA 91125, USA}
\author{Y.~Nakato}
\affiliation{Department of Physics, Stanford University, Stanford, CA 94305, USA}
\author{T.~Namikawa}
\affiliation{Kavli Institute for the Physics and Mathematics of the Universe (WPI), UTIAS, The~University~of~Tokyo, Kashiwa, Chiba 277-8583, Japan}
\author{H.~T.~Nguyen}
\affiliation{Jet Propulsion Laboratory, Pasadena, CA 91109, USA}
\author{R.~O'Brient}
\affiliation{Department of Physics, California Institute of Technology, Pasadena, CA 91125, USA}
\affiliation{Jet Propulsion Laboratory, Pasadena, CA 91109, USA}
\author[0000-0002-1343-2684]{R.~W.~Ogburn~IV}
\affiliation{Department of Physics, Stanford University, Stanford, CA 94305, USA}
\affiliation{Kavli Institute for Particle Astrophysics and Cosmology, SLAC National Accelerator Laboratory, 2575 Sand Hill Rd, Menlo Park, CA 94025, USA}
\author{S.~Palladino}
\affiliation{Department of Physics, University of Cincinnati, Cincinnati, OH 45221, USA}
\author[0000-0002-4436-4215]{M.~A.~Petroff}
\affiliation{Center for Astrophysics, Harvard \& Smithsonian, Cambridge, MA 02138, USA}
\author{T.~Prouve}
\affiliation{Service des Basses Temp\'{e}ratures, Commissariat \`{a} l'Energie Atomique, 38054 Grenoble, France}
\author[0000-0003-3983-6668]{C.~Pryke}
\affiliation{School of Physics and Astronomy, University of Minnesota, Minneapolis, MN 55455, USA}
\affiliation{Minnesota Institute for Astrophysics, University of Minnesota, Minneapolis, MN 55455, USA}
\author{B.~Racine}
\affiliation{Center for Astrophysics, Harvard \& Smithsonian, Cambridge, MA 02138, USA}
\affiliation{Aix-Marseille  Universit\'{e},  CNRS/IN2P3,  CPPM,  F-13288 Marseille,  France}
\author{C.~D.~Reintsema}
\affiliation{National Institute of Standards and Technology, Boulder, CO 80305, USA}
\author{S.~Richter}
\affiliation{Center for Astrophysics, Harvard \& Smithsonian, Cambridge, MA 02138, USA}
\author{A.~Schillaci}
\affiliation{Department of Physics, California Institute of Technology, Pasadena, CA 91125, USA}
\author{R.~Schwarz}
\affiliation{School of Physics and Astronomy, University of Minnesota, Minneapolis, MN 55455, USA}
\author{B.~L.~Schmitt}
\affiliation{Center for Astrophysics, Harvard \& Smithsonian, Cambridge, MA 02138, USA}
\author{C.~D.~Sheehy}
\affiliation{Physics Department, Brookhaven National Laboratory, Upton, NY 11973, USA}
\author[0000-0001-7387-0881]{B.~Singari}
\affiliation{Minnesota Institute for Astrophysics, University of Minnesota, Minneapolis, MN 55455, USA}
\author{A.~Soliman}
\affiliation{Department of Physics, California Institute of Technology, Pasadena, CA 91125, USA}
\author{T.~St.~Germaine}
\affiliation{Center for Astrophysics, Harvard \& Smithsonian, Cambridge, MA 02138, USA}
\affiliation{Department of Physics, Harvard University, Cambridge, MA 02138, USA}
\author{B.~Steinbach}
\affiliation{Department of Physics, California Institute of Technology, Pasadena, CA 91125, USA}
\author{R.~V.~Sudiwala}
\affiliation{School of Physics and Astronomy, Cardiff University, Cardiff, CF24 3AA, United Kingdom}
\author{G.~P.~Teply}
\affiliation{Department of Physics, California Institute of Technology, Pasadena, CA 91125, USA}
\author{K.~L.~Thompson}
\affiliation{Department of Physics, Stanford University, Stanford, CA 94305, USA}
\affiliation{Kavli Institute for Particle Astrophysics and Cosmology, SLAC National Accelerator Laboratory, 2575 Sand Hill Rd, Menlo Park, CA 94025, USA}
\author{J.~E.~Tolan}
\affiliation{Department of Physics, Stanford University, Stanford, CA 94305, USA}
\author[0000-0002-1851-3918]{C.~Tucker}
\affiliation{School of Physics and Astronomy, Cardiff University, Cardiff, CF24 3AA, United Kingdom}
\author{A.~D.~Turner}
\affiliation{Jet Propulsion Laboratory, Pasadena, CA 91109, USA}
\author{C.~Umilt\`{a}}
\affiliation{Department of Physics, University of Cincinnati, Cincinnati, OH 45221, USA}
\affiliation{Department of Physics, University of Illinois at Urbana-Champaign, Urbana, IL 61801, USA}
\author[0000-0002-3942-1609]{C.~Verg\`{e}s}
\affiliation{Center for Astrophysics, Harvard \& Smithsonian, Cambridge, MA 02138, USA}
\author{A.~G.~Vieregg}
\affiliation{Department of Physics, Enrico Fermi Institute, University of Chicago, Chicago, IL 60637, USA}
\affiliation{Kavli Institute for Cosmological Physics, University of Chicago, Chicago, IL 60637, USA}
\author[0000-0002-8232-7343]{A.~Wandui}
\affiliation{Department of Physics, California Institute of Technology, Pasadena, CA 91125, USA}
\author{A.~C.~Weber}
\affiliation{Jet Propulsion Laboratory, Pasadena, CA 91109, USA}
\author{D.~V.~Wiebe}
\affiliation{Department of Physics and Astronomy, University of British Columbia, Vancouver, British Columbia, V6T 1Z1, Canada}
\author[0000-0002-6452-4693]{J.~Willmert}
\affiliation{School of Physics and Astronomy, University of Minnesota, Minneapolis, MN 55455, USA}
\author{C.~L.~Wong}
\affiliation{Center for Astrophysics, Harvard \& Smithsonian, Cambridge, MA 02138, USA}
\affiliation{Department of Physics, Harvard University, Cambridge, MA 02138, USA}
\author[0000-0001-5411-6920]{W.~L.~K.~Wu}
\affiliation{Kavli Institute for Particle Astrophysics and Cosmology, SLAC National Accelerator Laboratory, 2575 Sand Hill Rd, Menlo Park, CA 94025, USA}
\author{H.~Yang}
\affiliation{Department of Physics, Stanford University, Stanford, CA 94305, USA}
\author{K.~W.~Yoon}
\affiliation{Department of Physics, Stanford University, Stanford, CA 94305, USA}
\affiliation{Kavli Institute for Particle Astrophysics and Cosmology, SLAC National Accelerator Laboratory, 2575 Sand Hill Rd, Menlo Park, CA 94025, USA}
\author{E.~Young}
\affiliation{Department of Physics, Stanford University, Stanford, CA 94305, USA}
\affiliation{Kavli Institute for Particle Astrophysics and Cosmology, SLAC National Accelerator Laboratory, 2575 Sand Hill Rd, Menlo Park, CA 94025, USA}
\author[0000-0002-8542-232X]{C.~Yu}
\affiliation{Department of Physics, Stanford University, Stanford, CA 94305, USA}
\author[0000-0001-6924-9072]{L.~Zeng}
\affiliation{Center for Astrophysics, Harvard \& Smithsonian, Cambridge, MA 02138, USA}
\author[0000-0001-8288-5823]{C.~Zhang}
\affiliation{Department of Physics, California Institute of Technology, Pasadena, CA 91125, USA}
\author{S.~Zhang}
\affiliation{Department of Physics, California Institute of Technology, Pasadena, CA 91125, USA}

\collaboration{95}{(BICEP/Keck Collaboration)}

\begin{abstract}
We characterize Galactic dust filaments by correlating BICEP/Keck and Planck data with polarization templates based on neutral hydrogen (H~\textsc{i}) observations. Dust polarization is important for both our understanding of astrophysical processes in the interstellar medium (ISM) and the search for primordial gravitational waves in the cosmic microwave background~(CMB). In the diffuse ISM, H~\textsc{i} is strongly correlated with the dust and partly organized into filaments that are aligned with the local magnetic field. We analyze the deep BICEP/Keck data at 95, 150, and 220 GHz, over the low-column-density region of sky where BICEP/Keck has set the best limits on primordial gravitational waves. We separate the H~\textsc{i} emission into distinct velocity components and detect dust polarization correlated with the local Galactic H~\textsc{i} but not with the H~\textsc{i} associated with Magellanic Stream~\textsc{i}. We present a robust, multifrequency detection of polarized dust emission correlated with the filamentary H~\textsc{i} morphology template down to 95~GHz. For assessing its utility for foreground cleaning, we report that the H~\textsc{i} morphology template correlates in $B$ modes at a~$\sim$10-65$\%$ level over the multipole range $20~<~\ell~<~200$ with the BICEP/Keck maps, which contain contributions from dust, CMB, and noise components. We measure the spectral index of the filamentary dust component spectral energy distribution to be $\beta = 1.54 \pm 0.13$. We find no evidence for decorrelation in this region between the filaments and the rest of the dust field or from the inclusion of dust associated with the intermediate velocity H~\textsc{i}. Finally, we explore the morphological parameter space in the H~\textsc{i}-based filamentary model.
\end{abstract}

\keywords{Interstellar dust (836) --- Interstellar filaments (842) --- Neutral hydrogen clouds (1099) --- Cosmic microwave background radiation (322) --- Interstellar magnetic fields (845) --- Interstellar medium (847) --- Interstellar atomic gas (833) --- Galaxy magnetic fields (604) --- Milky Way magnetic fields (1057) --- Magnetic fields (994) --- Interstellar phases (850)}

\section{Introduction} \label{sec:intro}
An accurate characterization of polarized dust emission is important for understanding different astrophysical phenomena in the interstellar medium~(ISM) and studying the polarization of the cosmic microwave background~(CMB). The short axes of aspherical rotating dust grains are preferentially aligned with the local magnetic field. This causes their thermal emission to be linearly polarized \citep{1975duun.book..155P}. Polarized dust emission is the dominant polarized CMB foreground at frequencies greater than approximately 70~GHz and at large scales \citep{2016A&A...594A..10P}. Characterizing and removing the dust contribution to CMB polarization measurements allows us to look for an excess signal generated by primordial gravitational waves, parameterized by the tensor-to-scalar ratio $r$, in order to constrain primordial gravitational waves \citep{Kamionkowski_1997,Seljak_1997,Seljak__1997}.

Galactic neutral hydrogen (H~\textsc{i}) gas has several advantages for tracing properties of the dust polarization. H~\textsc{i} is strongly correlated with dust throughout the diffuse ISM \citep{1996A&A...312..256B,2017ApJ...846...38L}. The dust and H~\textsc{i} are organized into filamentary structures \citep{2015PhRvL.115x1302C,2014A&A...571A..11P}. H~\textsc{i} filaments are well aligned with the plane-of-sky magnetic field orientation \citep{2014ApJ...789...82C,2015PhRvL.115x1302C}. Moreover, since the H~\textsc{i} measurements are spectroscopic, they provide~3D (position, position, and velocity) information about the H~\textsc{i} emission, where velocity is inferred from the Doppler-shifted frequency of the 21 cm line. They are also independent from the broadband thermal dust millimeter-wave and far infrared emission observations, and therefore, do not contain correlated systematics. Finally, H~\textsc{i} measurements are not contaminated by the cosmic infrared background \citep[CIB;][]{2019ApJ...870..120C}. These advantages allow us to exploit cross correlations between the data collected by CMB experiments and H~\textsc{i} surveys to better understand and characterize diffuse dust polarization. \citet{2019ApJ...887..136C} developed a formalism for modeling the linear polarization structure of Galactic dust emission solely from H~\textsc{i} intensity measurements. They have shown that these H~\textsc{i} morphology templates correlate at the $\sim$60\% ($\sim$50\%) level in $E$~modes ($B$~modes) with Planck data at 353~GHz at multipole $\ell=50$ over the high-Galactic latitude sky, and the correlation decays roughly monotonically to zero at around multipole moment $\ell\approx1000$.

The BICEP2 and Keck Array CMB experiments target a $\sim400~{\rm deg}^2$ patch of high-Galactic latitude sky \citep[][hereafter BK18]{2021PhRvL.127o1301A}. The instantaneous field of view of BICEP3 is larger and targets a $\sim600~{\rm deg}^2$ patch, which encompasses that of BICEP2 and Keck Array \citep{2022ApJ...927...77A}. These patches were chosen to have relatively little dust emission in intensity \citep{1999ApJ...524..867F}. In this paper, we use BICEP/Keck maps using all data taken up to and including the 2018 observing season, the data set known as ``BK18." These instruments have $\sim$30$\%$ fractional bandwidths and have achieved great depths at different frequencies. The polarization maps at 95, 150, and 220~GHz reach depths of 2.8, 2.8, and 8.8~$\mu {\rm K}_{\rm CMB}$~arcmin respectively \citep{2021PhRvL.127o1301A}. The signal-to-noise on polarized dust emission of the 220~GHz maps exceeds that of Planck at 353~GHz in the BICEP/Keck region \citep{2021PhRvL.127o1301A}. These data thus present an excellent opportunity to study the structure of the diffuse, magnetic ISM. Furthermore, this well-characterized region of sky will also be observed by future CMB experiments like CMB-S4 \citep{Abazajian_2022}. In this paper, we make use of cross correlations of BK18 data with H~\textsc{i} morphology maps. Because the H~\textsc{i} morphology templates are defined solely from the morphology of linear H~\textsc{i} structures, we refer to the component of the real dust field that is correlated with these templates as filamentary.

A motivation for using H~\textsc{i} to study dust in the BICEP/Keck region is its promise as a tracer of the 3D structure of the magnetic ISM \citep{2018ApJ...857L..10C, 2019ApJ...887..136C}. A differently oriented magnetic field along the line of sight will give rise to different dust polarization angles along that line of sight \citep{2015MNRAS.451L..90T}. If this dust is described by different spectral energy distributions (SEDs) in different locations along that sightline, the measured dust polarization angle will be frequency-dependent. This is referred to as line-of-sight frequency decorrelation. Frequency decorrelation can also arise due to spatial variations of the dust SED in the plane of the sky, producing frequency-dependent variations in the dust polarization pattern. Decorrelation causes maps of dust emission at different frequencies to differ by more than just a multiplicative factor, complicating the ability to use dust maps at one frequency to constrain the dust emission at another frequency. The decorrelation parameter,~$\Delta_{\rm d}$, defined as the ratio of the cross-spectrum between maps at 217 and 353~GHz to the geometric mean of the corresponding autospectra, is currently constrained to~$\Delta_{\rm d} > 0.98$ (68\% C.L.) in the BICEP/Keck region \citep{2021PhRvL.127o1301A}. Therefore, we currently have no indication of dust decorrelation in this region. However, there is evidence for frequency decorrelation in data, either associated with superpositions of independent line-of-sight emission \citep{2021A&A...647A..16P} or, at large scales, with spatial variations in the dust-polarization SED \citep{2022arXiv220607671R}. \citet{2021A&A...647A..16P} measure evidence for line-of-sight frequency decorrelation. They make a statistically significant detection of a stronger frequency-dependent change of the polarization angle along lines of sight which intercept multiple dust clouds with different magnetic field orientations. Therefore, it is interesting to isolate and separately characterize the distinct H~\textsc{i} velocity components along the line of sight in the region observed by BICEP2, BICEP3, and the Keck Array instruments to look for evidence for this effect. Additionally, we look for evidence of decorrelation due to any variation in the polarized dust SED between dust filaments, identified by the H~\textsc{i} morphology model and generally associated with the cold neutral medium \citep{Clark:2019,2020A&A...639A..26K}, and the rest of the dust column.

In this paper, we perform cross correlations between the Stokes parameter maps of the H~\textsc{i} morphology template and BICEP/Keck and Planck data and measure the statistical significance of the correlation as a function of frequency, instrument, and H~\textsc{i} velocity component in the BICEP/Keck region. To clarify, the H~\textsc{i}-based Stokes parameter maps are based on H~\textsc{i} morphology and not on H~\textsc{i} polarization. The cross correlations allow us to pick out the filamentary dust signal from the overall dust signal measured by BICEP/Keck and Planck in that region. We use our formalism to compare the sensitivities of Planck and BICEP/Keck in that region, to tune the H~\textsc{i} morphology template, and to search for frequency decorrelation. We also measure the SED of the dust correlated with H~\textsc{i} filaments. Knowledge of the dust SED is essential for CMB studies \citep{2017MNRAS.472.1195C,2018ApJ...853..127H} and for providing constraints for physical models of dust composition \citep[e.g.][]{2022ApJ...929..166H}.

This paper is organized as follows. We introduce the data used in this work in \secref{sec:data}. In \secref{sec:ana}, we introduce the methodology to estimate the statistical significance of the detection and to measure the filamentary dust SED. In \secref{sec:clouds}, we present a method for separating the different velocity components in the BICEP/Keck regions using H~\textsc{i} velocity information. Our results are presented and discussed in \secref{sec:results}. We then conclude with a summary and outlook in \secref{sec:conc}.

\section{Data} \label{sec:data}
\subsection{Millimeter-wave Polarization \label{subsec:mmwave}}
In this paper, we use BICEP3 data at 95 GHz from 2016 to 2018, BICEP2 data at 150 GHz from 2010 to 2012, and Keck Array data at 150 and 220 GHz from 2012 to 2018 \citep{2021PhRvL.127o1301A}. We also use the Planck NPIPE processed maps at 143, 217, and 353 GHz \citep{p2020}. These are a subset of the maps we used in BK18 to set the most stringent upper limits on the tensor-to-scalar ratio,~$r$. We do not consider the lower-frequency maps from CMB experiments, i.e. the 23 and 33 GHz bands of Wilkinson Microwave Anisotropy Probe (WMAP) and the 30 and 44 GHz bands of Planck, since we expect a negligible emission contribution from dust in those channels.

In \secref{subsec:tf}, we use the Planck 70\% sky fraction Galactic plane mask\footnote{Available for download at \url{http://pla.esac.esa.int}
(HFI\_Mask\_GalPlane-apo0\_2048\_R2.00.fits)} \citep{2015A&A...576A.104P} for calculating a transfer function for the H~\textsc{i} morphology template.

\subsection{Neutral Hydrogen Emission \label{subsec:hi}}
The H\textsc{i}4PI spectroscopic survey is the highest-resolution full-sky H~\textsc{i} survey to date \citep{2016A&A...594A.116H}. It has an angular resolution of $16'.2$, a spectral resolution of 1.49~km~${\rm s}^{-1}$, and a velocity-bin separation of 1.29~km~${\rm s}^{-1}$, achieved by merging data from the Effelsberg-Bonn H~\textsc{i} Survey \citep[EBHIS;][]{2016A&A...585A..41W} and the Parkes Galactic All-Sky Survey \citep[GASS;][]{2009ApJS..181..398M}. We start out with the velocity channels in the range $-120~{\rm km~s}^{-1}~<~v_{\rm lsr}~<~230$~km~s$^{-1}$, because the H\textsc{i}4PI maps are noise dominated in the BICEP/Keck region outside that range. We use these data to form H~\textsc{i} morphology templates as described in \secref{subsec:convRHT}.

\section{Methodology} \label{sec:ana}
\subsection{Convolutional Rolling Hough Transform \label{subsec:convRHT}}
\citet{2019ApJ...887..136C} used the Rolling Hough Transform \citep[RHT;][]{2014ApJ...789...82C,2020ascl.soft03005C} on the H\textsc{i}4PI data to construct 3D (position, position, and velocity) Stokes parameter maps. The mapping defined from H~\textsc{i} emission to properties of the dust polarization is based on several observational facts, including that the H~\textsc{i} column density correlates well with dust in the diffuse ISM \citep[][]{1996A&A...312..256B,2017ApJ...846...38L}. Also, H~\textsc{i} gas contains substantial linear structures that are preferentially aligned with the plane-of-sky component of the local magnetic field \citep{2015PhRvL.115x1302C}. Therefore, the dust polarization angle is taken to be orthogonal to these filaments. \citet{2019ApJ...887..136C} have shown that these maps, integrated over the velocity dimension \citep{2018ApJ...857L..10C}, are highly correlated with the Planck maps of the polarized dust emission at 353~GHz. 

While recent work over large regions of high-Galactic latitude sky (not focused on the BICEP/Keck region) has shown that there may be a small aggregate misalignment between the filaments and the Planck-measured magnetic field orientation \citep[][]{Huffenberger_2020,2021ApJ...919...53C}, the misalignment angle is only~$\sim2^\circ-5^\circ$ and incorporating it increases the correlation by only an additive~$\sim$0.1\%-0.5\% \citep{Cukierman}.

The first step of the RHT algorithm involves subtracting a smoothed version of the map from the original unsmoothed map. This is known as an unsharp mask and is used to remove the diffuse, large-scale H~\textsc{i} emission. This introduces a free parameter that sets the scale of the Gaussian smoothing filter. We refer to this parameter as the smoothing radius~($\theta_{\rm FWHM}$). The second step is to quantize the pixels into a bit mask, where the pixels are turned into zeros and ones based on their sign in the unsharp-masked data. The third step is to apply the Hough transform \citep{osti_4746348} on a circular window of a given diameter centered on each pixel. The window diameter~($D_W$) is the second parameter of this algorithm. The fourth step is to retain only values above a certain threshold fraction of the window diameter, where the threshold fraction~($Z$) is the third and last parameter. Refer to \citet{2014ApJ...789...82C} for further details. 

The RHT quantifies the intensity of linear structures as a function of orientation \citep{2014ApJ...789...82C}. Following \citet{2019ApJ...887..136C}, we use the RHT output to construct Stokes~$Q$ and $U$ polarization maps, weighted by the H~\textsc{i} intensity. Together, the RHT parameters~($\theta_{\rm FWHM}$,~$D_W$,~$Z$) determine what H~\textsc{i} filament morphologies most influence the H~\textsc{i} morphology template. It is thus of interest to explore the RHT parameter space and cross correlate different H~\textsc{i} morphology templates with the real dust polarization measurements, in order to determine what H~\textsc{i} morphologies are most predictive of the true polarized dust emission. Exploring the parameter space of the original RHT implementation was found to be computationally expensive, limited by the application of the Hough transform to each circular window of data. Other applications have used a convolutional implementation of the Hough transform \citep[e.g.,][]{CHT}. By rewriting the Hough transform step of the RHT as a series of convolutions, one for each orientation bin, we achieved a $\sim 35\times$ speedup in the RHT algorithm runtime. This convolutional implementation is made public via the RHT GitHub repository \citep{2020ascl.soft03005C}. In this work, we apply the convolutional RHT to the H\textsc{i}4PI data in the BICEP/Keck region to construct a 3D H~\textsc{i} morphology template.

\begin{figure}[t!]
\plotone{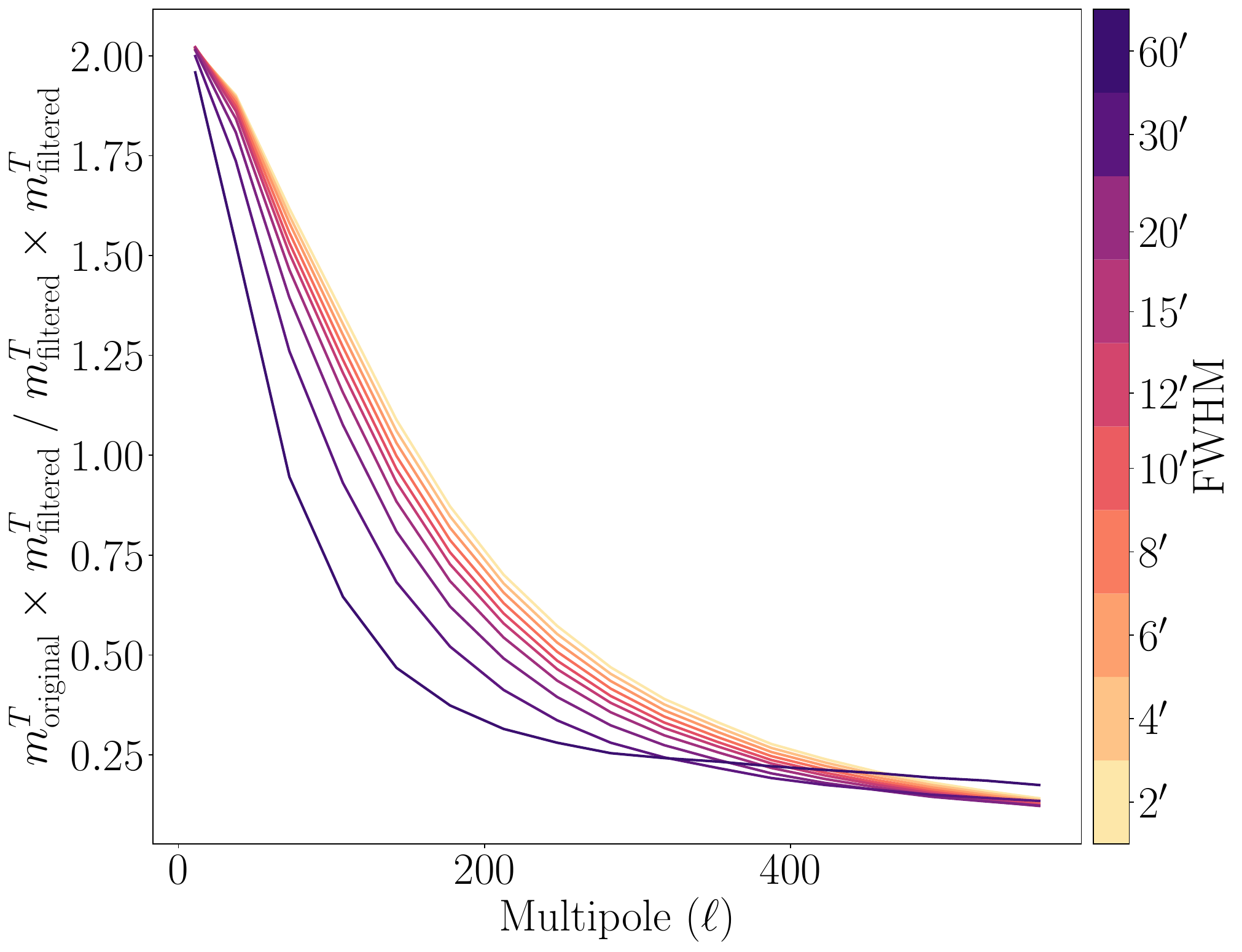}
\caption{The RHT algorithm multipole-dependent unitless transfer function defined in \eqref{eq:tf} for different Gaussian smoothing FWHM values, computed on the Planck 70\% sky fraction Galactic plane mask. \label{fig:TF}}
\end{figure}

\subsection{RHT Transfer Function \label{subsec:tf}}
The H~\textsc{i} morphology templates have different mode structures than the dust maps. As described in \secref{subsec:convRHT}, one of the first steps of the RHT algorithm is an unsharp mask. This filter emphasizes small-scale features. For instance, the $E$- and $B$-mode autospectra of the templates constructed with the same RHT parameters as those used in \citet{2019ApJ...887..136C} peak in the multipole range~$300~<~\ell~<~500$ and~$150~<~\ell~<~350$, respectively. We denote these spectra by~$D_\ell^{{\rm HI} \times {\rm HI}} = \ell (\ell + 1) C_\ell^{{\rm HI} \times {\rm HI}} / (2 \pi)$, where~$C_\ell^{m_1 \times m_2}$ is the cross spectrum bandpower between two maps, $m_1$ and $m_2$, in the multipole bin $\ell$. Correlation ratios are insensitive to this mode structure because the relative weightings of different multipole bins are normalized out of the calculation. Although the H~\textsc{i} morphology template itself shows a suppression of large-scale modes, the correlation with millimeter-wave polarization is strongest at large scales. The statistical tests defined in this paper, however, are based on cross spectra rather than correlation ratios. We form cross spectra between the data collected by CMB experiments and the H~\textsc{i} morphology template defined in \secref{subsec:convRHT}, and we denote these spectra by~$D_\ell^{{\rm data} \times {\rm HI}}$. 

We cannot make a direct comparison between~$D_\ell^{{\rm data} \times {\rm HI}}$ and~$D_\ell^{{\rm HI} \times {\rm HI}}$, because they are not, in general, proportional to each other. As in \citet{Cukierman}, we model this effect as a multipole-dependent transfer function that describes the representation of the H~\textsc{i} morphology template in the measured dust polarization. We denote the transfer function by~$t_\ell$. The goal in constructing~$t_\ell$ is for~$D_\ell^{{\rm data} \times {\rm HI}}$ to be approximately proportional to~$t_\ell D_\ell^{{\rm HI} \times {\rm HI}}$. In our statistical tests, we will compare the former cross spectra to the latter multipole-filtered autospectra.

The aim in introducing the transfer function $t_\ell$ is to boost large-scale modes relative to small-scale modes in order to enhance the sensitivity of our statistical tests. The best estimate of~$t_\ell$ would come from~$D_\ell^{{\rm data} \times {\rm HI}}/D_\ell^{{\rm HI} \times {\rm HI}}$ \citep[as in][]{Cukierman}, but this would lead to a fitting function ($t_\ell D_\ell^{{\rm HI} \times {\rm HI}}$) which is partly defined by the data itself. To avoid those complications, we use an ansatz based on the unsharp-mask filter, which produces most of the multipole distortion we wish to correct. This multipole correction is an ansatz and not a model of the true underlying reality. We use it in the same manner as a matched filter, i.e., to increase the sensitivity of our signal search by looking for a particular pattern rather than simply looking for deviations from zero. A discrepancy between the ansatz and the true reality would simply degrade our sensitivity.

To calculate this transfer function based on the unsharp-mask filter, we apply the following steps to the H~\textsc{i} emission maps at each velocity channel:
\begin{enumerate}
 \item{Smooth the original H~\textsc{i} intensity map with a Gaussian filter of a specific FWHM.}
 \item{Subtract the smoothed map from the original map.}
 \item{Quantize into a bit mask, i.e. set pixels with values~$> 0$ to 1 and pixels with values~$< 0$ to 0.}
 \item{Multiply the bit mask by the original map.}
\end{enumerate}
These are the subset of the steps in the RHT algorithm that most substantially restrict the range of spatial scales of the H~\textsc{i} emission that contributes to the measured H~\textsc{i} orientation. The subsequent steps, the Hough transform and thresholding, introduce further scale-dependent effects that effectively set the minimum length of a detected linear feature.

We sum the filtered velocity channel maps and call this the \textit{filtered} map. We refer to the velocity-integrated H~\textsc{i} intensity as the \textit{original} map. Because we do not expect this transfer function to vary dramatically over the sky, we use the Planck 70\% sky fraction Galactic plane mask \citep{2015A&A...576A.104P} as opposed to the BICEP/Keck mask for calculating the transfer function in order to obtain higher signal-to-noise and to capture the filtering effect better over the lower multipole bins. We define the transfer function as
\eq{t_\ell = \frac{C_\ell^{\mathrm{original} \times \mathrm{filtered}}}{C_\ell^{\mathrm{filtered} \times \mathrm{filtered}}}. \label{eq:tf}}
We consider the standard 9~bins in the angular multipole range $20 < \ell < 335$ that we use in BICEP/Keck analyses. Note that the only free parameter of the RHT algorithm that is used in this filtering is the Gaussian smoothing radius $\theta_{\rm FWHM}$. In \figref{fig:TF}, we plot this transfer function for the list of $\theta_{\rm FWHM}$ values we analyze. This is applied to the H~\textsc{i}-correlated component of the simulation in harmonic space. For the rest of this analysis, we present our results with the use of this transfer function. Repeating the analysis without the transfer function produces qualitatively similar results (see Appendix~\ref{app:variations}).

In the next subsection, we will describe a simulation construction that contains a component based on the H~\textsc{i} morphology template. We incorporate the multipole correction in the simulation construction such that~$D_\ell^{{\rm data}\times{\rm HI}}$ is approximately proportional to~$t_\ell~D_\ell^{{\rm HI}\times{\rm HI}}$. An explicit prescription is provided in the next section.

\subsection{BICEP/Keck and Planck Simulations Including Filamentary Dust\label{subsec:sims}}
We construct a set of mock realizations of the sky as observed by the BICEP/Keck and Planck instruments in order to check for biases and estimate uncertainties in the statistical tests introduced in subsequent sections. The baseline dust model in BICEP/Keck analyses is a statistically isotropic Gaussian-dust (GD) field and is our null-hypothesis dust model in this analysis. We call this model GD. It is uncorrelated with the H~\textsc{i} morphology template. Simulations of this model are created as random Gaussian realizations with a power spectrum defined by its amplitude~$A_{\rm d,353} = 3.75~\mu{\rm K}_{\rm CMB}^2$ at multipole moment $\ell = 80$ and frequency~$\nu=353$~GHz. The power spectrum scales spatially as a power law with index $\alpha_{\rm d} = -0.4$ in mutipole \citep{2021PhRvL.127o1301A}. In addition to the baseline dust model, we introduce a second component of filamentary dust that is perfectly correlated with the H~\textsc{i} morphology template (HI). This is one realization based on real H~\textsc{i} morphology that is added to 499 realizations of GD. 

We modify the H~\textsc{i}-correlated component in harmonic space according to the transfer function defined in \secref{subsec:tf} and inverse transform back to map space. We denote the multipole-filtered version of the H~\textsc{i} morphology template with a tilde ($\tilde{{\rm HI}}$). It is important to note that the transfer function introduced in \secref{subsec:tf} is a phenomenological ansatz rather than a model for the true multipole dependence of the H~\textsc{i}-correlated component of dust polarization. We use this ansatz as a fitting function in \secref{subsec:statTests} in order to improve the sensitivity of our search for H~\textsc{i}-correlated dust polarization, but the ansatz is likely only a rough approximation to the underlying reality. Indeed, we find moderate discrepancies between the measured H~\textsc{i}-dust cross-spectra and the fitting-function ansatz (see \figref{fig:BBcomps}). Furthermore, there is no guarantee that the H~\textsc{i} morphology template should appear in the dust field with a correction that depends only on multipole. If this assumption is made, however, a better estimate of the transfer function can be achieved by appealing to the H~\textsc{i}-dust cross-spectra themselves, which is how a similar transfer function is constructed in \citet{Cukierman}. As mentioned in \secref{subsec:tf}, however, we wish for our fitting function to be independent of the data to which we are fitting, so we prefer, for the purposes of statistical tests, the ansatz based on the unsharp-mask filtering. For the purposes of constructing mock-sky realizations, it may be superior to use the data-based transfer function in order to keep the mean cross-spectrum bandpowers identical to those of the real data. For computational simplicity, however, we use only the transfer function of \secref{subsec:tf} for all of the results in this paper. When our mock-sky realizations are used with a nonzero H~\textsc{i}-correlated component, we will only be interested in the variance of our fitting parameters. In the limit of relatively small perturbations, the variance in the fitting parameters is independent of the mean, so we expect our variance estimates to be reliable in spite of the discrepancy between the measured bandpowers and the mean of the simulated bandpowers.

The full dust field at frequency $\nu$ is modeled as
\begin{eqnarray}
m\p{\mathrm{d}}_\nu(\unitVec{n},a,k,\beta_{\rm HI}) &\equiv& a \cdot f_\nu(\beta_{\rm GD}) \cdot m\p{\mathrm{GD}} (\unitVec{n})\\ & &+ k \cdot f_\nu(\beta_{\rm HI}) \cdot m\p{\tilde{\mathrm{HI}}}(\unitVec{n}) , \nonumber
\end{eqnarray}
where $m(\unitVec{n})$ represents a Stokes $Q$ or $U$ map, and $a$, $k$, and $\beta_{\rm HI}$ are free parameters. The amplitude $a$ is unitless, and $k$ acts as both an amplitude and a unit conversion factor with units $\mu$K$_{\rm CMB}$~/~K~km~s$^{-1}$ because $m\p{\mathrm{GD}} (\unitVec{n})$ has units~$\mu$K$_{\rm CMB}$ and $m\p{\tilde{\mathrm{HI}}}(\unitVec{n})$ has units~K~km~s$^{-1}$. We use a modified blackbody scaling law~$f_\nu$ with a fixed temperature, $T =  19.6\,{\rm K}$, and variable frequency spectral index $\beta$ \citep[e.g.,][]{2014}. The exact choice of dust temperature is of little consequence for our measurements, because we are measuring at frequencies far below the thermal peak. We fix $\beta_{\rm GD}=1.6$ in our fiducial model, which is close to the value inferred from data. The exact value does not affect the results because the observables we use in the statistical tests in \secref{subsec:statTests} are cross correlations with the H~\textsc{i} morphology template, and the GD and HI components are uncorrelated. In the baseline tensor-to-scalar ratio analysis of BICEP/Keck, we model the dust on the level of cross-frequency $B$-mode power spectra. In this context, the full dust model of this paper would manifest itself as
\begin{eqnarray}
D_\ell^{\nu_1 \times \nu_2} &=& a^2 A_d f_{\nu_1}(\beta_{\rm GD}) f_{\nu_2}(\beta_{\rm GD}) \left(\frac{\ell}{80}\right)^{\alpha_d}\\ & &+ k^2 f_{\nu_1}(\beta_{\rm HI}) f_{\nu_2}(\beta_{\rm HI}) D_\ell^{\tilde{{\rm HI}}\times\tilde{{\rm HI}}} \nonumber.
\end{eqnarray}

We recover the standard dust model used in BICEP/Keck analyses (the null hypothesis) by setting~$a = 1$, and~$k = 0$. This hybrid model of GD and $\tilde{\rm{HI}}$ is continuously related to the GD null hypothesis because the null hypothesis is nested within the hybrid model. We also consider a variation of this model in Appendix~\ref{app:variations}, replacing $f_\nu$ with a power-law frequency scaling, and find that it does not affect the results, as expected in the Rayleigh-Jeans limit.

In this paper, we limit our analysis to the $\sim400~{\rm deg}^2$ region mapped by BICEP2 and Keck Array, centered at R.A. $0^{\rm h}$, decl. -57$^{\circ}$.5 (hereafter the BICEP/Keck region). On this small region, we use a flat-sky approximation.

We convolve the H~\textsc{i} morphology template with instrument-specific beams of different sizes. We also apply the instrument-specific observation matrices used in the BICEP/Keck cosmological analyses, $\mathbf{R}_\nu$, capturing the linear filtering of $Q$ and $U$ maps, which includes data selection, polynomial filtering, scan-synchronous signal subtraction, weighting, binning into map pixels, and deprojection of leaked temperature signal \citep{2016ApJ...825...66B}. We define
\eq{\tilde{m}_\nu\p{\tilde{\mathrm{HI}}}(\unitVec{n}) = \mathbf{R}_\nu(m\p{\tilde{\mathrm{HI}}}(\unitVec{n})),}
where $\tilde{m}_\nu\p{\tilde{\mathrm{HI}}}$ is the reobserved H~\textsc{i}-correlated component of the simulation.

Following standard procedure in BICEP/Keck analyses, we add lensed-$\Lambda$CDM ($\Lambda$CDM) and noise (n) components to the dust realizations. Refer to BK18 for more details of these simulations. For Planck, we use the official noise simulations provided in the NPIPE data release \citep{p2020}.

The model for our total, observed map at frequency~$\nu$ then becomes
\begin{eqnarray}
\label{eq:filtered model}
\tilde{m}_\nu(\unitVec{n},a,k,\beta_{\rm HI}) &=& \tilde{m}_\nu\p{\mathrm{\Lambda CDM}}(\unitVec{n}) + \tilde{m}\p{\rm n}_\nu(\unitVec{n})\\ & &+ a \cdot f_\nu(\beta_{\rm GD}) \cdot \tilde{m}_\nu\p{\mathrm{GD}} (\unitVec{n}) \nonumber
\\ & &+ k \cdot f_\nu(\beta_{\rm HI}) \cdot \tilde{m}_\nu\p{\tilde{\mathrm{HI}}}(\unitVec{n}). \nonumber 
\end{eqnarray}

We also purify the maps at each observing frequency with a matrix operation such that the resulting $B$~modes are cleaned of leakage from the much brighter $E$~modes \citep{2016ApJ...825...66B}. We then apodize the maps with an inverse noise variance weighting, Fourier transform them, and rotate them from a~$Q$/$U$ to an $E$/$B$ basis.

We refer to the real BICEP/Keck and Planck maps described in \secref{subsec:mmwave} as~$\tilde{m}\p{\mathrm{real}}_\nu(\unitVec{n})$.

\subsection{Cross Spectra \label{subsec:xspecs}}
The statistical tests defined in this paper are based on power spectra calculated using the standard power spectrum estimator of BICEP/Keck analyses as we described in BK18. We consider 9~bins in the angular multipole range~$20~<~\ell~<~335$ and compute both $EE$ and $BB$ autospectra. We then exploit the linearity of \eqref{eq:filtered model} to decompose the full cross spectrum with the H~\textsc{i} morphology template and calculate the binned bandpower expectation values as
\begin{eqnarray}
\label{eq:obs}
D_\ell^{{\rm data} \times {\rm HI}}(a,k,\beta_{\rm HI}) &=&D_\ell^{\mathrm{\Lambda CDM} \times {\rm HI}} + D_\ell^{{\rm n} \times {\rm HI}} \\& &+ a \cdot f_\nu(\beta_{\rm GD}) \cdot D_\ell^{{\rm GD} \times {\rm HI}} \nonumber\\& &+ k \cdot f_\nu(\beta_{\rm HI}) \cdot D_\ell^{\tilde{{\rm HI}} \times {\rm HI}}.\nonumber
\end{eqnarray}

We concatenate the 9 bandpowers of \eqref{eq:obs} for a selection of frequencies over $EE$ only, $BB$ only, or $EE$ and $BB$ into $\mathbf{D}(a,k,\beta_{\rm HI})$.
The vector $\mathbf{D}(a,k,\beta_{\rm HI})$ contains the observables from which we construct the covariance matrix in \secref{subsec:covmats} and our statistical tests in \secref{subsec:statTests}. We similarly define the vector of cross spectra of the real data with the H~\textsc{i} morphology template for a selection of frequencies over $EE$ only, $BB$ only, or $EE$ and $BB$ as~$\mathbf{D}\p{\mathrm{real}}$.

\subsection{Covariance Matrices \label{subsec:covmats}}
To construct covariance matrices, we start with 499 realizations of \eqref{eq:obs} of the fiducial model, which coincides with the null-hypothesis model used in the standard BICEP/Keck analyses, i.e. $a~=~1$ and $k~=~0$. In the covariance matrix construction, we neglect variances of the H~\textsc{i}-correlated dust component because we expect any uncertainty from the H~\textsc{i} data itself to be subdominant.

There are nonnegligible covariances between neighboring multipole bins and, because the lensed-$\Lambda$CDM and dust fields are broadband, between frequency channels. Therefore, we construct a covariance matrix of the form,
\begin{eqnarray}
\mathbf{M} &\equiv& \frac{N}{N-1} \langle (\mathbf{D}(1,0,0) - \overline{\mathbf{D}}(1,0,0))\\ & & \otimes (\mathbf{D}(1,0,0) - \overline{\mathbf{D}}(1,0,0)) \rangle_{\rm rlz} , \nonumber
\label{eq:cov matrix def}
\end{eqnarray}
where~$\overline{\mathbf{D}}$ is the mean of the vector of spectra over realizations, $N$ is the number of realizations, $\otimes$ is an outer product, and $\langle\rangle_{\rm rlz}$ is a mean over realizations.

For the statistical test discussed in the next subsection, we use different combinations of the 95, 150, and 220 GHz channels of BICEP/Keck and the 143, 217, and 353 GHz channels of Planck. We use 9 bandpowers per spectrum and separately consider only $B$~modes, only $E$~modes, and $E$~and $B$~modes simultaneously. We condition the covariance matrix by forcing some entries to zero \citep[e.g.,][]{2022MNRAS.515..229B}. We allow covariances between neighboring multipole bins and between any two frequencies (not just neighboring frequencies), and neglect the correlations between $E$~and $B$~modes in our covariance matrix construction.

\subsection{Statistical Tests \label{subsec:statTests}}
In this subsection, we define the statistical tests that are used in \secref{sec:results} of this paper.
\subsubsection{$\chi^2$ Likelihood \label{subsubsec:chisqlike}}
We approximate the cross spectra defined in \secref{subsec:xspecs} between the simulations for our total, observed, maps and the H~\textsc{i} morphology templates as Gaussian distributed, so the natural choice for a test statistic to fit our model is 
\begin{eqnarray}
\label{eq:chi2 def}
\chi^2(a,k,\beta_{\rm HI}) &\equiv& \parens{ \mathbf{D}\p{\mathrm{real}} - \overline{\mathbf{D}}(a,k,\beta_{\rm HI}) }^{\rm T} \\& & \mathbf{M}^{-1} \parens{ \mathbf{D}\p{\mathrm{real}} - \overline{\mathbf{D}}(a,k,\beta_{\rm HI}) } , \nonumber 
\end{eqnarray}
where, again, $\overline{\mathbf{D}}$ is the mean of the vector of spectra over 499 realizations.

To calibrate this test statistic through simulations, we input an ensemble of realizations from \eqref{eq:obs} with~$a~=~1$ and~$k~=~0$ in place of $\mathbf{D}\p{\mathrm{real}}$. We fit the model by minimizing \eqref{eq:chi2 def} with respect to the three model parameters~$a$, $k$, and~$\beta_{\rm HI}$. We form the test statistic
\eq{ \hat{\chi}^2 \equiv \chi^2(\hat{a},\hat{k},\hat{\beta}_{\rm HI}) , \label{eq:chi2min} }
where $\hat{a}$, $\hat{k}$, and~$\hat{\beta}_{\rm HI}$ are the model parameters that minimize \eqref{eq:chi2 def} (e.g., \secref{subsec:sed}).

Because our observables are cross-spectra between the H~\textsc{i} morphology template and the dust polarization, we expect little sensitivity to the GD amplitude~$a$. We retain~$a$ as a fitting parameter, however, so that our null hypothesis ($a$ = 1, $k$ = 0) is nested within the full fitting function. This will allow us to form the more sensitive $\Delta\chi^2$ test statistic in \secref{subsubsec:DeltaChi2}. Another approach to this analysis could have been to fit for $k$ and $\beta$ only and to report the statistical significance in terms of the number of standard deviations of $\hat{k}$ from 0. However, we rely on the $\chi^2$ distribution to estimate statistical significance.

When the data are drawn from the null-hypothesis model, the minimized test statistic~$\hat{\chi}^2$ is expected to be~$\chi^2$~distributed with~$n-3$ degrees of freedom, where~$n$ is the number of observables used. For the cases where we only use one frequency band to estimate each band's contribution to the statistical significance of the detection, $k$ and $\beta_{\rm HI}$ are degenerate. We therefore fit $\hat{k} f_\nu(\hat{\beta}_{\rm HI})$ as one value. In those cases, there are only 2 effective parameters, $a$ and $k f_\nu(\beta_{\rm HI})$, and $\hat{\chi}^2$ is $\chi^2$ distributed with~$n-2$ degrees of freedom.

We also use a Markov Chain Monte Carlo (MCMC) method to fully explore this parameter space and provide insight into the correlations and degeneracies between these parameters. We use noninformative uniform distributions for the priors, [-50,~50], [0,~5], and~[0.8,~2.4], on $a$, $k$, $\beta_{\rm HI}$, respectively. The range is large for $a$ because the GD cross spectra with the H~\textsc{i} morphology template have no constraining power for~$a$. Using the $\chi^2$ likelihood defined in \eqref{eq:chi2 def}, we sample the posterior distributions using the Metropolis-Hastings algorithm implemented in the \textsc{cobaya} MCMC Python package \citep{2019ascl.soft10019T,2021JCAP...05..057T}. 

\subsubsection{$\Delta\chi^2$ Detection Significance Metric \label{subsubsec:DeltaChi2}}
We form a $\Delta\chi^2$ statistic for measuring the statistical significance of detecting the H~\textsc{i} morphology template. We compare~$\hat{\chi}^2$ from \eqref{eq:chi2min} to a model in which $a$, the amplitude of GD, is allowed to vary but for which $k = 0$. This comparison isolates the influence of the H~\textsc{i}-related degrees of freedom.

We form the test statistic
\eq{ \chi^2_\mathrm{GD}(a) \equiv \chi^2(a,0,0) \label{eq:chi2Gd 1template def} }
and we minimize with respect to~$a$ to obtain
\eq{ \hat{\chi}^2_\mathrm{GD} \equiv \chi^2_\mathrm{GD}(\hat{a}\p{\mathrm{GD}}) , \label{eq:chi2Gd min def} }
where $\hat{a}\p{\mathrm{GD}}$~is the best-fit value for the model with GD only. The test statistic~$\hat{\chi}^2_\mathrm{GD}$ is expected to be~$\chi^2$~distributed with $n -1$ degrees of freedom when the data are drawn from the null-hypothesis distribution.

We test for the added benefit of the H~\textsc{i}-correlated component with the test statistic
\eq{ \Delta\chi^2 = \hat{\chi}_\mathrm{GD}^2 - \hat{\chi}^2 , \label{eq:DeltaChi2 def} }
which is expected to be $\chi^2$~distributed with $2$~degrees of freedom when the data are drawn from the null-hypothesis distribution. If only a single frequency band is used, then $\Delta\chi^2$ is expected
to be $\chi^2$ distributed with only 1 degree of freedom.

The statistical significance of the correlation between the data and the H~\textsc{i} morphology template can be estimated from~$\Delta\chi^2$. The ensemble of $\Delta\chi^2$ measurements from the null-hypothesis simulations matches a $\chi^2$~distribution with the given number of degrees of freedom. This allows us to calculate a p-value or a probability to exceed (PTE) as PTE~=~1~-~CDF, where CDF is the cumulative distribution function of the ensemble up to the $\Delta\chi^2$ value we get from the data. We convert the PTE to an equivalent Gaussian deviate to present the significance as a number of standard deviations from the mean. The reported significances, however, are less reliable $\gtrapprox3\sigma$, where there are no $\Delta\chi^2$ measurements from the null-hypothesis simulations.

\begin{figure*}[t!]
\plotone{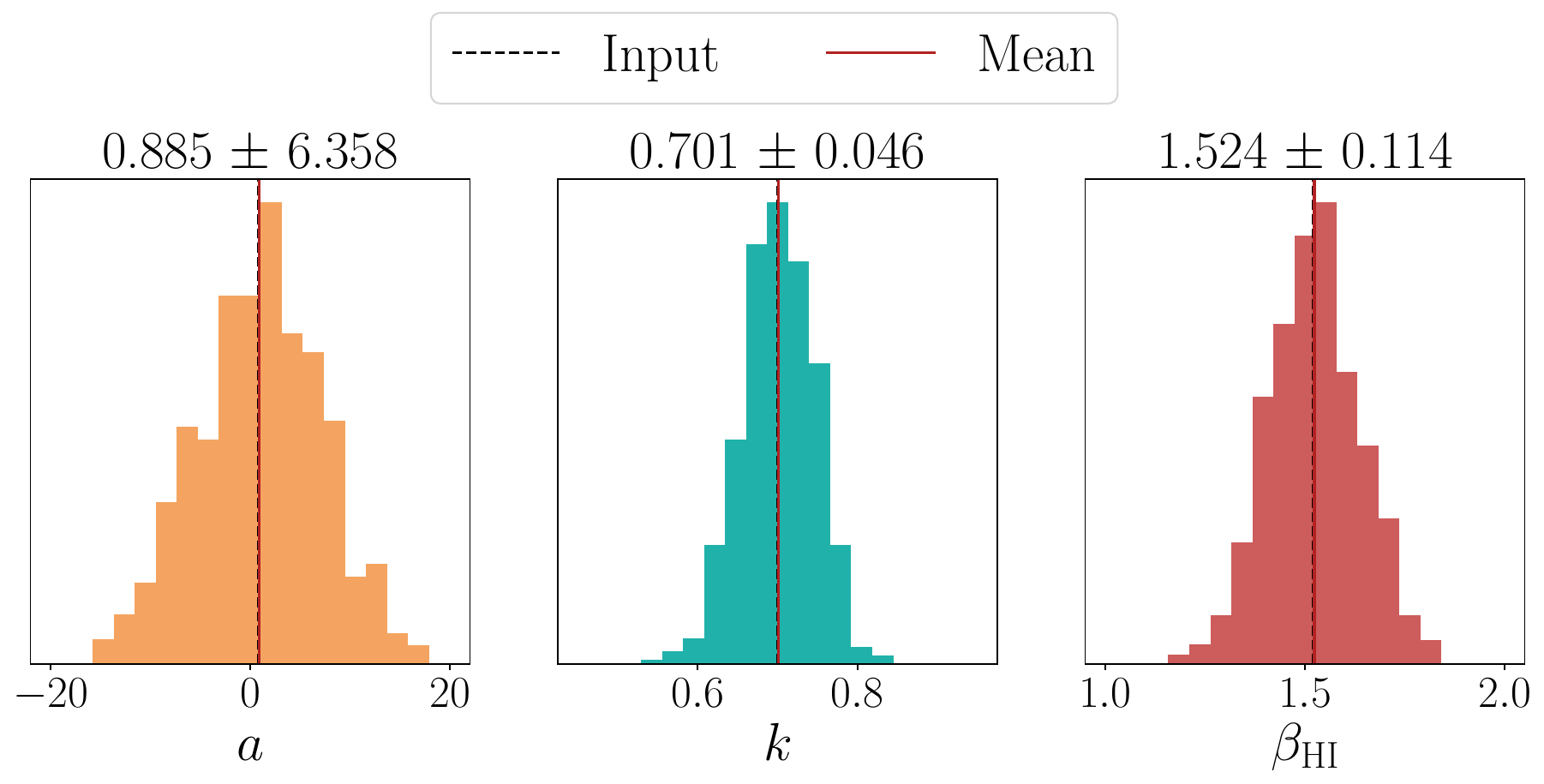}
\caption{
Distributions of the best-fit values using $E$~and $B$~modes for 499 realizations of lensed-$\Lambda$CDM, noise, and Gaussian dust, added to the H~\textsc{i} morphology template with fixed input values $a=0.9$, $k=0.7$, and $\beta_{\rm HI}=1.52$ that match the fit from the real data. The parameters $a$ and $\beta_{\rm HI}$ are unitless, and $k$ has units $\mu$K$_{\rm CMB}$~/~K~km~s$^{-1}$. These known input values are plotted as dashed black vertical lines. The means of the distributions of the best-fit values are plotted as solid red vertical lines. The mean and standard deviation of each of the distributions are quoted above. 
\label{fig:unc}}
\end{figure*}
\begin{figure*}[t!]
\plotone{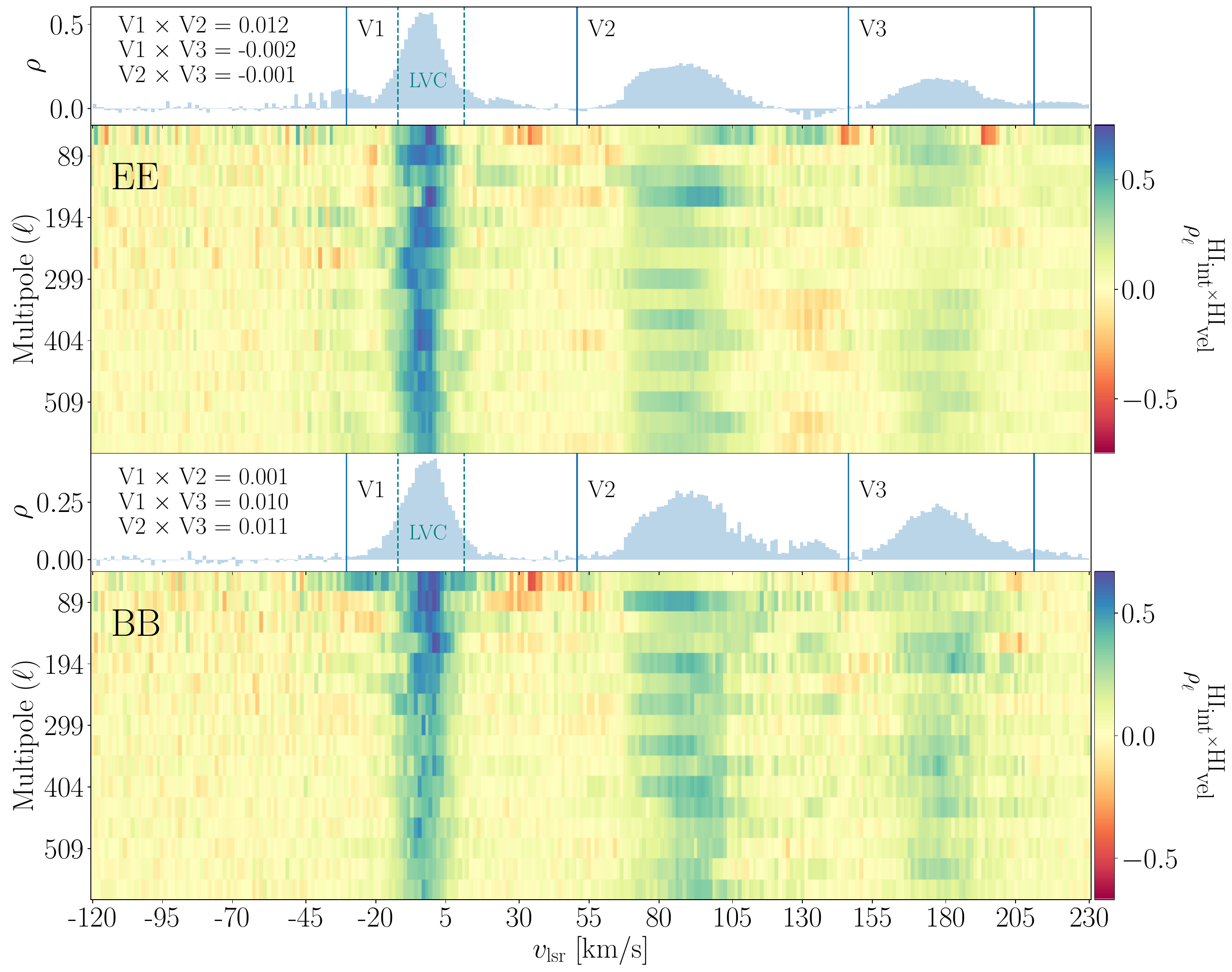}
\caption{
$EE$ (top) and $BB$ (bottom) correlation ratio of the integrated H~\textsc{i} morphology template with individual H~\textsc{i} morphology templates for the H\textsc{i}4PI velocity channels across multipoles $37~<\ell<~579$. The 1D plots on top show the broadband correlation ratio calculated over one mutipole bin spanning the entire multipole range. It is separated into 3 velocity regions, V1, V2, and V3. The LVC boundaries as defined in \citet{2020ApJ...902..120P} are indicated with dashed vertical lines. The broadband correlation ratio between the different pair combinations of the 3 velocity components is printed on the left of each histogram. \label{fig:clouds}}
\end{figure*}

\begin{figure*}[t!]
\plotone{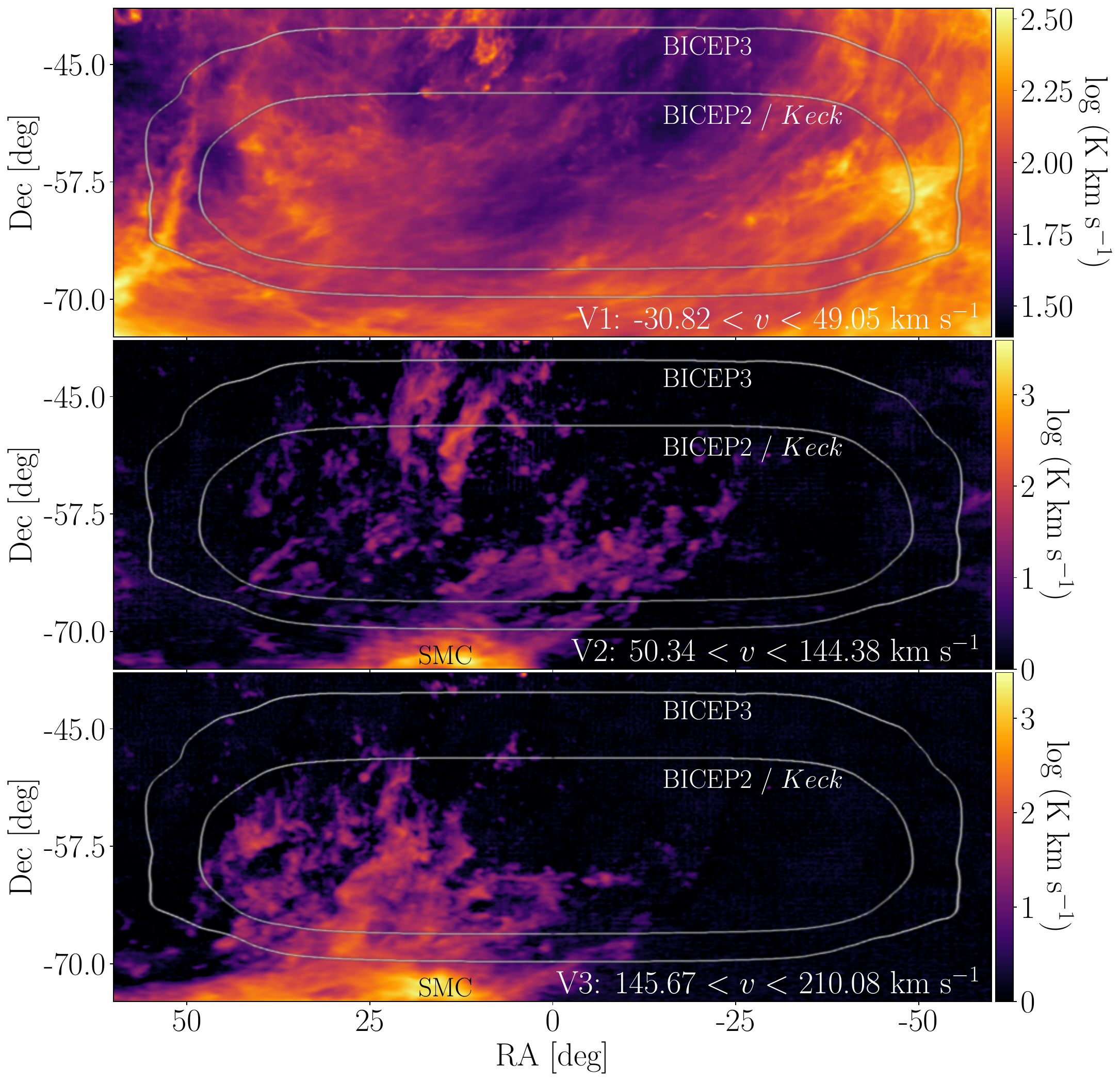}
\caption{Integrated H~\textsc{i} intensity maps over the 3 different velocity components defined in \figref{fig:clouds} in the BICEP/Keck region. The velocity boundaries for each component are printed on the bottom right of each map. The emission in V1 is dominated by the Milky Way, whereas the emission in V2 and V3 is dominated by Magellanic Stream~\textsc{i} \citep{2018MNRAS.474..289W}. The outlines of the BICEP3 and the BICEP2 and Keck Array observing fields are also plotted. The Small Magellanic Cloud (SMC) is indicated. \label{fig:maps}}
\end{figure*}

\subsection{Parameter Estimation \label{subsec:uncertainty}}
We perform a coverage test of our Bayesian model by computing the maximum-likelihood values of a simulation set of 499 realizations with fixed $a$, $k$, and $\beta_{\rm HI}$ values and compare their distributions to the posteriors obtained from real data. We use the best-fit results for $k$ and $\beta_{\rm HI}$ from the real data. We fit $a$, such that the autospectrum of the total dust field is equivalent to the GD autospectrum used in BICEP/Keck analyses, and the cross spectrum of the total dust field with the H~\textsc{i} morphology template is equivalent to the best-fit autospectrum of the H~\textsc{i} morphology template. We call this best-fit $\hat{\alpha}$ to distinguish it from the best-fit $\hat{a}$ we get from \secref{subsubsec:chisqlike}. We refer the reader to Appendix \ref{app:unc} for a detailed description of this fit.

We then repeat the statistical test defined in \secref{subsubsec:chisqlike}, replacing $\mathbf{D}\p{\rm real}$ with each of the cross spectra of these 499 realizations with the H~\textsc{i} morphology template, and get a distribution of 499 best-fit values for each parameter. Example distributions of the best-fit values from these realizations are shown in \figref{fig:unc}. The distributions shown here are from fitting $E$~and $B$~modes simultaneously using the 95, 150, and 220~GHz bands of BICEP/Keck and the 143, 217, and 353 GHz bands of Planck, conditioning the covariance matrix, and using a transfer function for the H~\textsc{i} morphology template with RHT parameters
\begin{equation}
\label{eq:bestparams}
D_W = 135',\,\, \theta_{\rm FWHM} = 4',\,\, {\rm and }\,\, Z = 0.75.
\end{equation}
This RHT parameter selection is motivated in \secref{subsec:rhttune} and is the fiducial set we use in the results of this paper unless otherwise mentioned. For these choices, the fixed input values used for constructing the simulation set are 0.9, 0.7, and 1.52 for $a$, $k$, and $\beta_{\rm HI}$, respectively. We find that our parameter estimation method is unbiased. The sample mean of $a$ is notably close to the input value relative to the standard error, but we checked the p-value and found it to be~4.1\%, which we deem to be small but acceptable. We conclude that our fits are unbiased, and we use the spread of the distributions for the~499 realizations to obtain an estimate of the parameter uncertainties. These are consistent with the uncertainties inferred from the marginalized posterior distributions in \secref{subsec:sed}, which are 
6.7, 0.050, and 0.13
for $a$, $k$, and $\beta_{\rm HI}$, respectively. The standard deviation for~$a$ is relatively large because the GD cross spectra with the H~\textsc{i} morphology template have no constraining power for $a$, and this parameter is marginalized over in our analysis.

\section{Velocity Decomposition} \label{sec:clouds}
At the high-Galactic latitudes considered here, there is no simple one-to-one mapping between the Galactic H~\textsc{i} emission's velocity along the line of sight and the distance to the H~\textsc{i} gas. However, the bulk velocity of clouds at various distances will often differ, resulting in distinct kinematic components in the H~\textsc{i} spectra. Utilizing the velocity dimension of the 3D H~\textsc{i} morphology Stokes parameter maps in the BICEP/Keck region, we can separate the different velocity components contributing the most to the polarization of the H~\textsc{i} morphology template along the line of sight.

We integrate the H~\textsc{i} morphology Stokes parameter maps in the BICEP/Keck region across the velocity dimension over the range $-120~{\rm km~s}^{-1}~<~v_{\rm lsr}~<~230$~km~s$^{-1}$ (see \secref{subsec:hi}) to form the maps $Q_{\rm int}$ and $U_{\rm int}$. This is analogous to the line-of-sight integration inherent in thermal dust emission measurements. We then correlate this integrated map with the maps for each velocity channel, H~\textsc{i}$_{\rm vel}$, using the correlation ratio defined as
\eq{ \rho_\ell^{X_{\rm int} \times X_{\rm vel}} = \frac{D_\ell^{X_{\rm int} \times X_{\rm vel}}}{\sqrt{D_\ell^{X_{\rm int} \times X_{\rm int}} \times D_\ell^{X_{\rm vel} \times X_{\rm vel}}}},
}
where $X$ denotes either the $E$ or $B$~modes of the H~\textsc{i} morphology templates, and $D_\ell$ is the cross spectra over multipole moment $\ell$. This metric quantifies the contribution of each velocity channel map to the polarization signal of the line-of-sight integrated template. We use the RHT parameters in \eqref{eq:bestparams} for this plot; though the results are qualitatively similar when varying those parameters.

We plot $\rho_\ell^{X_{\rm int} \times X_{\rm vel}}$ in \figref{fig:clouds}, where each column represents the correlation of each velocity channel map with the integrated map, and each row represents a multipole moment bin. We expect neighboring velocity channels to be correlated on physical grounds. Therefore, the consistent horizontal bands at each multipole bin in the 2D plots are due to the similarity between adjacent velocity channels.

We also calculate a broadband correlation coefficient that is binned into one multipole bin that spans the entire range ($37~<\ell<~579$) and plot it above the 2D plots in \figref{fig:clouds}. We clearly see distinct peaks in three different velocity ranges, which we refer to as V1, V2, and~V3. These peaks are in roughly the same locations as the peaks we see when plotting the H~\textsc{i} intensity as a function of velocity but have different relative amplitudes, with the second peak having a much lower amplitude in intensity than the third peak. We plot vertical lines to define roughly where the boundaries between those components are. As we will show in \secref{sec:results} and \tabref{tab:sigbound}, the exact boundaries do not affect the results, which are dominated by the velocity channels at the peaks.

The H~\textsc{i} line emission at high-Galactic latitudes is conventionally divided into low-velocity clouds (LVCs), intermediate-velocity clouds (IVCs), and high-velocity clouds
(HVCs) based on its radial velocity with respect to the local standard of rest ($v_{\rm lsr}$) or the Galactic standard of rest ($v_{\rm gsr}$), or on its deviation from a simple model of Galactic rotation (see, e.g., \citet{Putman_2012} for more details). The boundaries between these classes vary by tens of kilometers per second in the literature. For instance, \citet{2010ApJ...722.1685M}, \citet{1991A&A...250..499W}, and \citet{2001} define the boundary between LVCs and IVCs at~$| v_{\rm lsr} |~=~20, 30$, and~40~km~s$^{-1}$, respectively. \citet{2020ApJ...902..120P} propose $-12~{\rm km~s}^{-1}~<~v_{\rm lsr}~<~10$~km~s$^{-1}$ as the range for LVCs based on the first and 99th percentiles of the distribution of cloud velocities that pass a certain threshold in the H~\textsc{i} column density in the Northern and Southern Galactic Polar regions.

The boundary between IVCs and HVCs is usually taken to be at $| v_{\rm lsr} |~=~70$~km~s$^{-1}$ \citep{1986A&A...170...84W} or $90$~km~s$^{-1}$ \citep{Richter2005}. 
The boundaries for V1 defined here encompass the range of LVCs adopted by \citet{2020ApJ...902..120P} as shown in \figref{fig:clouds}. We limit the higher end of the IVC range to $| v_{\rm lsr} |~=~50$~km~s$^{-1}$ in the BICEP/Keck region such that V2, which is primarily associated with the Magellanic System \citep{2018MNRAS.474..289W}, is excluded. As already mentioned, the results are dominated by the velocity channels at the peaks, and the exact boundaries do not affect the results. 

Our interpretation of the peaks in \figref{fig:clouds} is that each corresponds to a substantial contribution of that velocity component to the integrated map. As a sanity check, however, we test whether the V2 and V3 peaks in the correlation with the integrated map are due to spurious correlations with each other or with V1 by calculating~$\rho^{X_{{\rm V}i} \times X_{{\rm V}j}}$, where $i$ and $j \in \{1,2,3 \mid i \neq j \}$. We report those values in \figref{fig:clouds} and find that the correlation is less than approximately~$1\%$.

We integrate the velocity channel maps in each range and plot the resulting H~\textsc{i} intensity maps in \figref{fig:maps} on a log color scale. V1 is dominated by H~\textsc{i} emission from the Galaxy, whereas V2 and V3 are dominated by H~\textsc{i} emission from Magellanic Stream~\textsc{i}, a stream of high-velocity gas associated with the Magellanic System \citep{2018MNRAS.474..289W}. The outlines of the BICEP3 and the BICEP2 and Keck Array observing fields are included in the figure to distinguish the H~\textsc{i} structure that lies inside and outside each of the observing fields. For consistency in our statistical tests defined in \secref{subsec:statTests}, we analyze the smaller field as mentioned in \secref{subsec:sims}. The bright emission in V2 and V3 directly below the BICEP3 observing field in decl. is from the Small Magellanic Cloud~(SMC).

\begin{table}
\centering
\begin{tabular}{c|c|c}
 & Default & Best \\
\tableline
$BB$ & 4.7 & 6.7\\
\tableline
$EE$ & 12.3 & 14.6\\
\tableline
$BB + EE$ & 12.9 & 16.1\\
\end{tabular}
\caption{Statistical significance of the detection of V1 in units of equivalent Gaussian standard deviations as defined in \secref{subsubsec:DeltaChi2} using the 95, 150, and 220 GHz bands of BICEP/Keck and the 353 GHz band of Planck. The column labeled ``best" uses the parameters $D_W = 135'$, $\theta_{\rm FWHM} = 4'$, and $Z = 0.75$, and the row labeled ``default" uses the parameters $D_W = 75'$, $\theta_{\rm FWHM} = 30'$, and $Z = 0.7$, which are used in \citet{2019ApJ...887..136C}.}
\label{tab:sigdefbest}
\end{table}

\begin{figure*}[t!]
\plotone{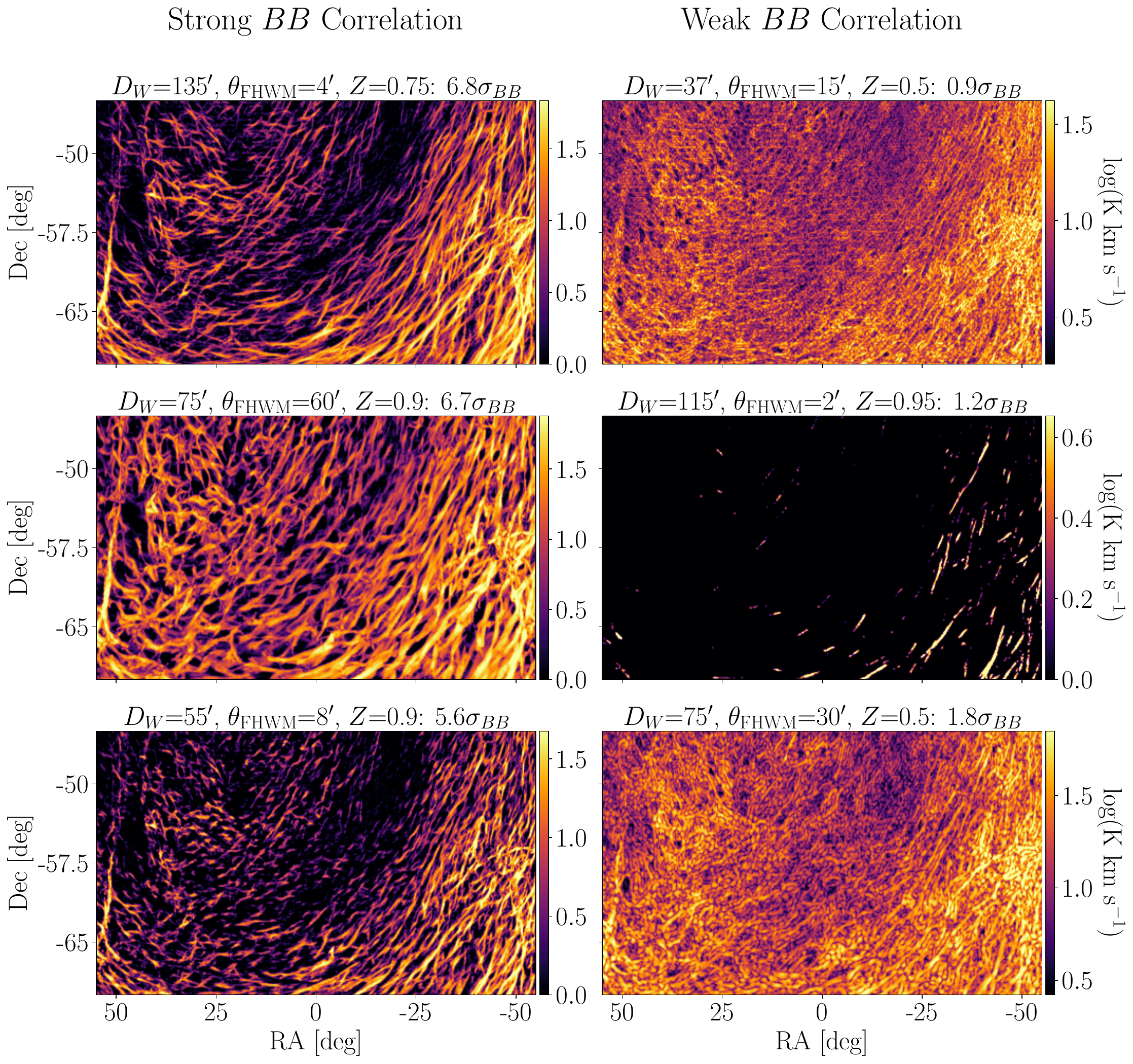}
\caption{
Polarized intensity maps of V1 in the BICEP/Keck region using RHT parameters that correlate $> 5 \sigma$ (left) and~$< 5 \sigma$ (right) in $B$~modes with BICEP/Keck and Planck data. Only the statistical significance in $B$~modes is quoted in the title of each of the maps, because all of the RHT parameters we tried correlate well ($> 5\sigma$) in $E$~modes. From top to bottom, the maps on the left have a 15.2$\sigma_{EE}$, 12.6$\sigma_{EE}$, and 14.9$\sigma_{EE}$ detection significances, and the maps on the right have a 6.3$\sigma_{EE}$, 8.6$\sigma_{EE}$, and~8.2$\sigma_{EE}$ detection significances.
\label{fig:RHTgvsb}}
\end{figure*}

\begin{figure*}[t!]
\plotone{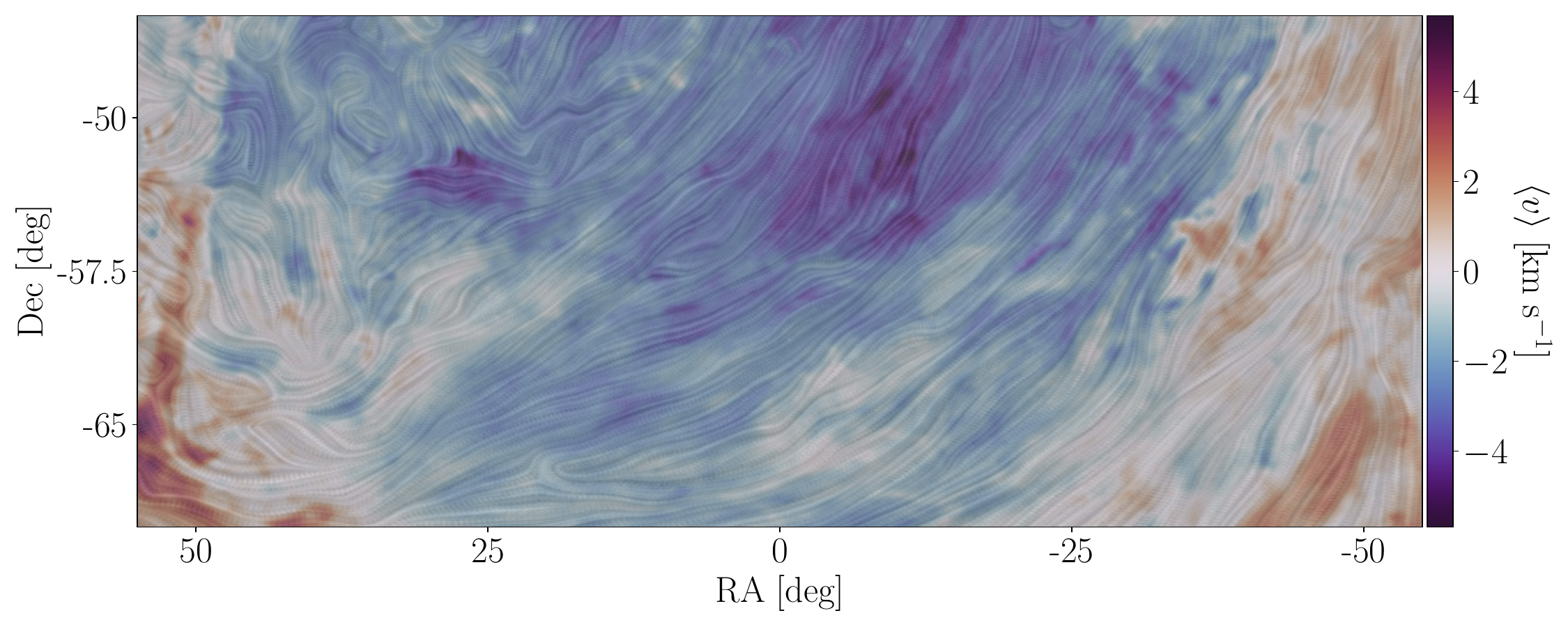}
\caption{Map of the first moment of the velocity distribution of the H~\textsc{i} structure in the BICEP/Keck region for $-12~{\rm km~s}^{-1}~<~v_{\rm lsr}~<~10$~km~s$^{-1}$, the velocity range most correlated with the polarized dust emission. The texture is a line integral convolution of the magnetic field orientation as inferred by the H~\textsc{i} filaments. \label{fig:lic}}
\end{figure*}

\section{Results and Discussion} \label{sec:results}
In this section, we tune the RHT parameters to increase the correlation between BICEP/Keck and Planck data with the H~\textsc{i} morphology template (\secref{subsec:rhttune}). Using the tuned parameters, we quantify the detection of filamentary dust polarization in the Galactic component of H~\textsc{i} (\secref{subsec:v1}). We look for evidence of frequency decorrelation in the BICEP/Keck region from the inclusion of the IVC component in the line-of-sight sum and between the filamentary dust component and the total dust component (\secref{subsec:sed}). We also quantify the contribution of each of the datasets used in this measurement (\secref{subsec:bkvp}). Finally, we look for a detection of filamentary dust polarization in the higher-velocity H~\textsc{i} components associated with Magellanic Stream~\textsc{i} (\secref{subsec:magstream}).

\subsection{Tuning and Improving the RHT Model \label{subsec:rhttune}}
Due to computational expense, the RHT parameter space has not been explored before in the context of building dust polarization templates. However, limiting the sky area to the BICEP/Keck region and speeding up the algorithm by $\sim 35\times$, as described in \secref{subsec:convRHT}, have allowed us to search the parameter space more efficiently. We evaluate the $\Delta\chi^2$ metric from \secref{subsubsec:DeltaChi2} in parallel on a grid of values spanning a reasonable range of interest in each of the RHT parameters. We consider $D_W =$ 37$'$, 55$'$, 75$'$, 95$'$, 115$'$, 135$'$, and 149$'$; $\theta_{\rm FWHM} =$ 2$'$, 4$'$, 6$'$, 8$'$, 10$'$, 12$'$, 15$'$, 30$'$, and 60$'$; and $Z=$ 0.5, 0.7, 0.75, 0.8, 0.85, 0.9, and 0.95.

We find that the RHT parameters that maximize the statistical significance of the detection among the ones we tried are $D_W = 135'$, $\theta_{\rm FWHM} = 4'$, and $Z = 0.75$. These parameters maximize the statistical significance when fitting the metric using $B$~modes only, $E$~modes only, and $E$~and $B$~modes simultaneously. We compare the results we get using these parameters to the results we get using the RHT parameters used in \citet{2019ApJ...887..136C} ($D_W = 75'$, $\theta_{\rm FWHM} = 30'$, and $Z = 0.7$) in \tabref{tab:sigdefbest}. Our results improve by $\sim$2$\sigma$ in $BB$ and in $EE$ and by $\sim$3$\sigma$ when $EE$ and $BB$ are combined.

We look for trends in the detection significance with BICEP/Keck and Planck data when varying each of the RHT parameters monotonically, but do not find any. Instead, we find that the correlation is robust for a wide range of parameter choices. The exceptions are at the extremes of the parameter space. We show examples of polarized intensity maps of the H~\textsc{i} morphology templates made with parameters that correlate well or poorly with the dust in $BB$ in \figref{fig:RHTgvsb}. The polarized intensity is defined as 
\eq{P = \sqrt{Q^2 + U^2}, \label{eq:polint}}
where $Q$ and $U$ are the Stokes parameters of the H~\textsc{i} morphology template. We quote the statistical significance of the detection in~$BB$ in the title of each panel. All of the variations we tried correlate well ($> 5 \sigma$) in $EE$, including the ones shown in \figref{fig:RHTgvsb} with their detection significances stated in the caption. Note that the examples that are weakly correlated with the dust in $BB$ either have a high $Z$ ($Z\gtrsim$~0.95) and $D_W \gg \theta_{\rm FWHM}$ or have a low $Z$ ($Z\lesssim$~0.5). While the significance is larger than 5$\sigma$ in $EE$ for the examples on the right, it is still fairly low by $E$-mode standards compared to the examples on the left with a lower correlation ratio.

The cases with a high $Z$ limit the RHT-detected linear structure to longer, more connected filaments, while lower $Z$ decomposes the H~\textsc{i} intensity into numerous shorter filaments. The choppiness of the filaments affects the predicted $B$-mode power more than it does the $E$-mode power because the $B$-mode structure of this template is affected by the finite extent of the filaments. Real-space maps of the $E$- and $B$-mode amplitudes support this intuition \citep{Huffenberger_2020}. The net signal arising from choppy, colinear filaments produces a constructive interference for $E$~modes but a destructive interference for $B$~modes. Also, because $\theta_{\rm FWHM}$ affects the largest spatial scales of the H~\textsc{i} emission and the product of the $D_W$ and the $Z$ parameters defines an effective lower limit on the length of the filaments, the combination of high $Z$ with $D_W \gg \theta_{\rm FWHM}$, such as the middle right panel of \figref{fig:RHTgvsb}, discards most of the structure in the map and is only sensitive to the most prominent filaments. The $B$-mode-correlated H~\textsc{i} structure is related to the overall distribution of filaments, such that annihilating all but a few substantially weakens the correlation with the dust $B$~modes.

We defer a more comprehensive interpretation of the RHT parameters and their implications to a future study. For now, we propose the parameters in \eqref{eq:bestparams} as the recommended ones when using the RHT in future analyses on H\textsc{i}4PI data for making dust polarization or magnetic field templates in the diffuse, high-Galactic latitude ISM. However, these parameters might be sensitive to the BICEP/Keck filtering or to the specific sky region. These effects will be explored in future work.

\subsection{Filamentary Polarization in the Local ISM \label{subsec:v1}}

Using the $\Delta\chi^2$ statistical test defined in \secref{subsubsec:DeltaChi2}, we find a significant correlation between the H~\textsc{i} morphology templates and the first velocity component, V1, as shown in \tabref{tab:sigdefbest}. These results are insensitive to covariance matrix conditioning, frequency scaling law, or use of a transfer function for the H~\textsc{i} morphology template as shown in Table \ref{apptab:sigdefbest}. 

Above a certain threshold in the column density of H~\textsc{i}, \citet{2020ApJ...902..120P} find an agreement between the Northern and Southern Galactic Polar regions in the first and 99th percentiles of the H~\textsc{i} cloud velocity distributions. They therefore use those percentiles to adopt the boundaries $-12~{\rm km~s}^{-1}~<~v_{\rm lsr}~<~10$~km~s$^{-1}$ between LVCs and IVCs. We use this range to visualize the first moment map of the velocity distribution of the H~\textsc{i} structure in the BICEP/Keck region in \figref{fig:lic}. That is, we plot the intensity-weighted mean velocity,
\eq{\langle v \rangle = \frac{\sum_v v \cdot I(v)}{\sum_v I(v)}, \label{eq:mom}}
to highlight the regions in the map where the emission is dominated by different velocities. This is the velocity range that exhibits the most substantial contribution to the dust-correlated template as we show in \secref{subsec:sed}. We perform a line integral convolution \citep{10.1145/166117.166151} on the H~\textsc{i} morphology $Q$ and $U$ maps in that velocity range, smoothed to the RHT window diameter scale, to visualize the magnetic field orientation inferred by the H~\textsc{i} filaments and overplot it as the texture in \figref{fig:lic}.

\subsection{Frequency Decorrelation and the Polarized Dust SED \label{subsec:sed}}
Dust components along the same line of sight with different polarization angles and SEDs give rise to a phenomenon called line-of-sight frequency decorrelation. We test for evidence of this phenomenon in the BICEP/Keck region between the LVC and IVC components and between the filamentary and total dust components.

LVCs and IVCs are known to contain dust \citep{1996A&A...312..256B,1998ApJ...507..507R,2011A&A...536A..24P}. The velocity range of V1 spans both LVCs and IVCs using the velocity boundaries defined in \citet{2020ApJ...902..120P}. These are the same boundaries that \citet{2021A&A...647A..16P} use in their analysis of line-of-sight frequency decorrelation in Planck data. \citet{2020ApJ...902..120P} use a Gaussian decomposition of the H~\textsc{i} emission profiles to estimate the number of distinct clouds along each sightline. While they show that most sightlines in the BICEP/Keck region are dominated by one LVC cloud on average, they do detect more than one cloud along some sightlines. \citet{2021A&A...647A..16P} detect line-of-sight frequency decorrelation in the sightlines that contain LVCs and IVCs with different polarization angles predicted by H~\textsc{i} morphology. While we know from \citet{2020ApJ...902..120P} that IVCs are not an important fraction of the H~\textsc{i} column in the BICEP/Keck region, we check whether that is also true in polarization, i.e., whether the polarization inferred from the H~\textsc{i} morphology templates in the IVC velocity range contributes significantly to the correlation with dust polarization. We find that the IVC emission integrated over the BICEP/Keck region is $\sim$25\% of the V1 column in intensity and $\sim$10\% of the V1 column in polarized intensity. \tabref{tab:sigbound} shows that the detection significance is not strongly changed by the inclusion of IVC-associated H~\textsc{i} morphology template in the line-of-sight sum, as expected on account of the amplitude ratios. The shifts in detection significance are $\lesssim 0.3 \sigma$ in all cases.

\begin{table}
\centering
\begin{tabular}{c|c|c}
& Range for LVCs + IVCs & Range for LVCs \\
\tableline
$BB$ & 6.7 & 6.8 \\
\tableline
$EE$ & 14.6 & 14.3\\
\tableline
$BB + EE$ & 16.1 & 16.1\\
\end{tabular}
\caption{Comparison of the statistical significance of a detection of the cross correlation with the dust polarization in units of equivalent Gaussian standard deviations when including the channels in the IVC velocity range in the line-of-sight sum. The RHT parameters from \eqref{eq:bestparams} are used here for the H~\textsc{i} morphology template with the 95, 150, and 220 GHz bands of BICEP/Keck and the 353 GHz band of Planck.}
\label{tab:sigbound}
\end{table}

\begin{figure*}[t!]
\plotone{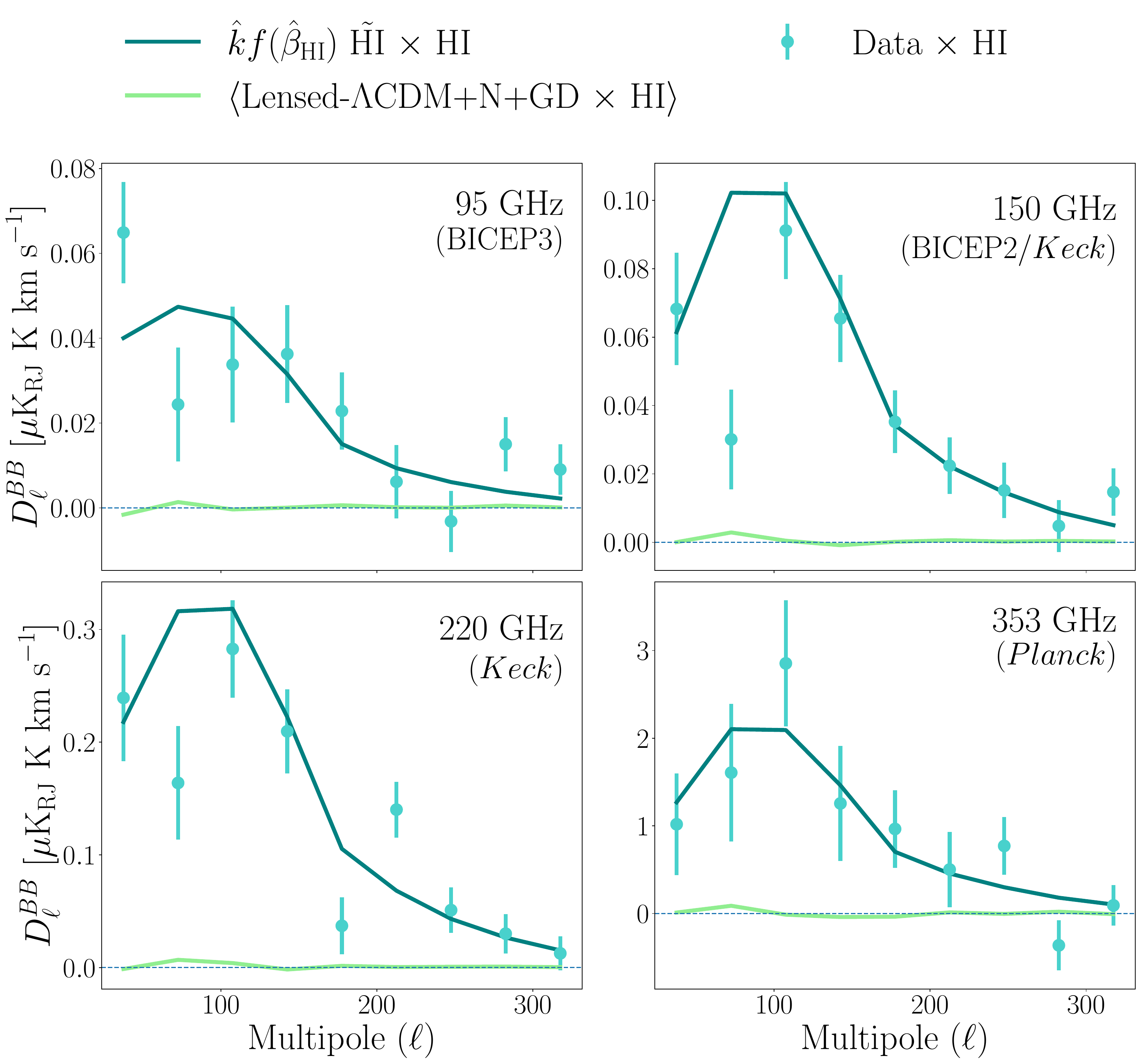}
\caption{The best-fit $BB$ observables used in the $\Delta\chi^2$ statistic defined in \secref{subsubsec:DeltaChi2} for the 95, 150, and 220 GHz bands of BICEP/Keck and the 353 GHz band of Planck. A modified blackbody frequency scaling, covariance matrix conditioning, and a transfer function for the H~\textsc{i} morphology template with the RHT parameters from \eqref{eq:bestparams} are used for the fit here. The cross spectrum between the real data and the H~\textsc{i} morphology template (light blue), the best-fit cross spectrum between the H~\textsc{i} morphology template and the modified H~\textsc{i}-correlated component of the simulation (dark blue), and the mean of the cross spectra between the H~\textsc{i} morphology template and the lensed-$\Lambda$CDM, noise, and Gaussian-dust components of the simulation (light green) are plotted.
\label{fig:BBcomps}}
\end{figure*}

Therefore, we do not have good reason to expect strong decorrelation from the IVC population in the BICEP/Keck region. However, there could be frequency decorrelation arising from different dusty regions along the line of sight that are all associated with gas within the LVC range. The kinematic substructure of the LVC H~\textsc{i} could in principle be used to further explore the 3D distribution and phase structure of the gas in this region, and its possible association with different contributions to the total dust SED.

Since the H~\textsc{i} morphology template is filamentary, the~$E$ and $B$ modes of this template are sourced by the same filaments \citep{Huffenberger_2020}, although variations in the 3D dust properties could still give rise to SED differences betwen~$E$ and $B$ modes \citep{vacher}. Minimizing the $\chi^2$ test statistic defined in \eqref{eq:chi2 def}, we fit $\beta$ using both~$E$~and~$B$~modes simultaneously. 

For the most sensitive measurement of $\beta_{\rm HI}$ in V1, we use both $E$~and $B$~modes, the best-fit RHT parameters from \eqref{eq:bestparams}, the 95, 150, and 220 GHz bands of BICEP/Keck, and the 143, 217, and 353 GHz bands of Planck. We condition the covariance matrix and use a transfer function for the H~\textsc{i} morphology template; though those choices do not substantially affect the result as shown in Appendix~\ref{app:variations}.

\begin{figure}[t!]
\plotone{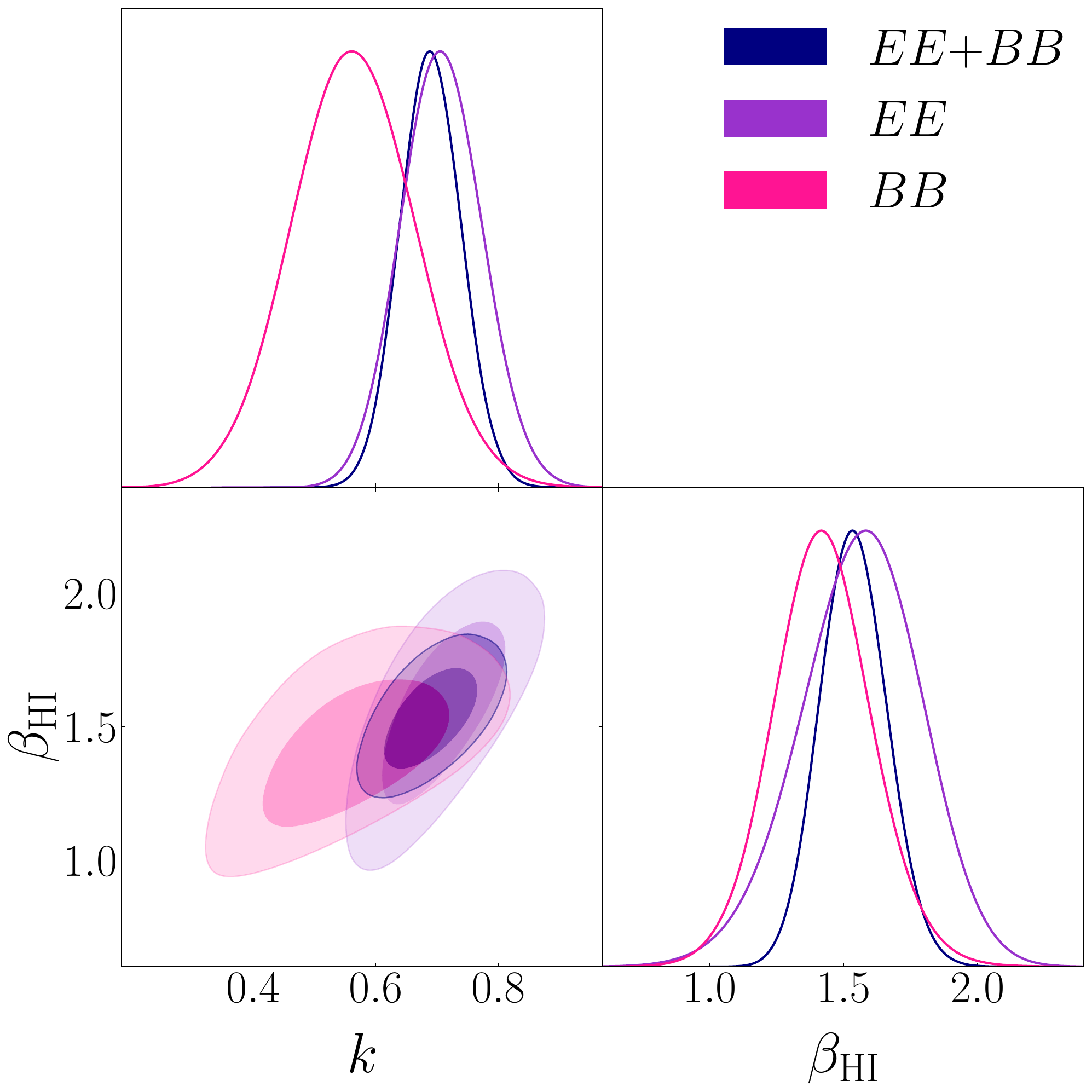}
\caption{
Posteriors of $k$ and $\beta_{\rm HI}$ fit using the Metropolis-Hastings algorithm on uniform priors and the $\chi^2$ likelihood of the cross spectra of the real data with the H~\textsc{i} morphology template. The parameter $a$ is marginalized over. The $E$~modes only (purple), $B$~modes only (pink), and simultaneous $E$~and $B$~modes (navy) posteriors are shown. The units for $k$ are $\mu$K$_{\rm CMB}$~/~K~km~s$^{-1}$ and $\beta_{\rm HI}$ is unitless.
\label{fig:corner}}
\end{figure}
\begin{figure}[t!]
\plotone{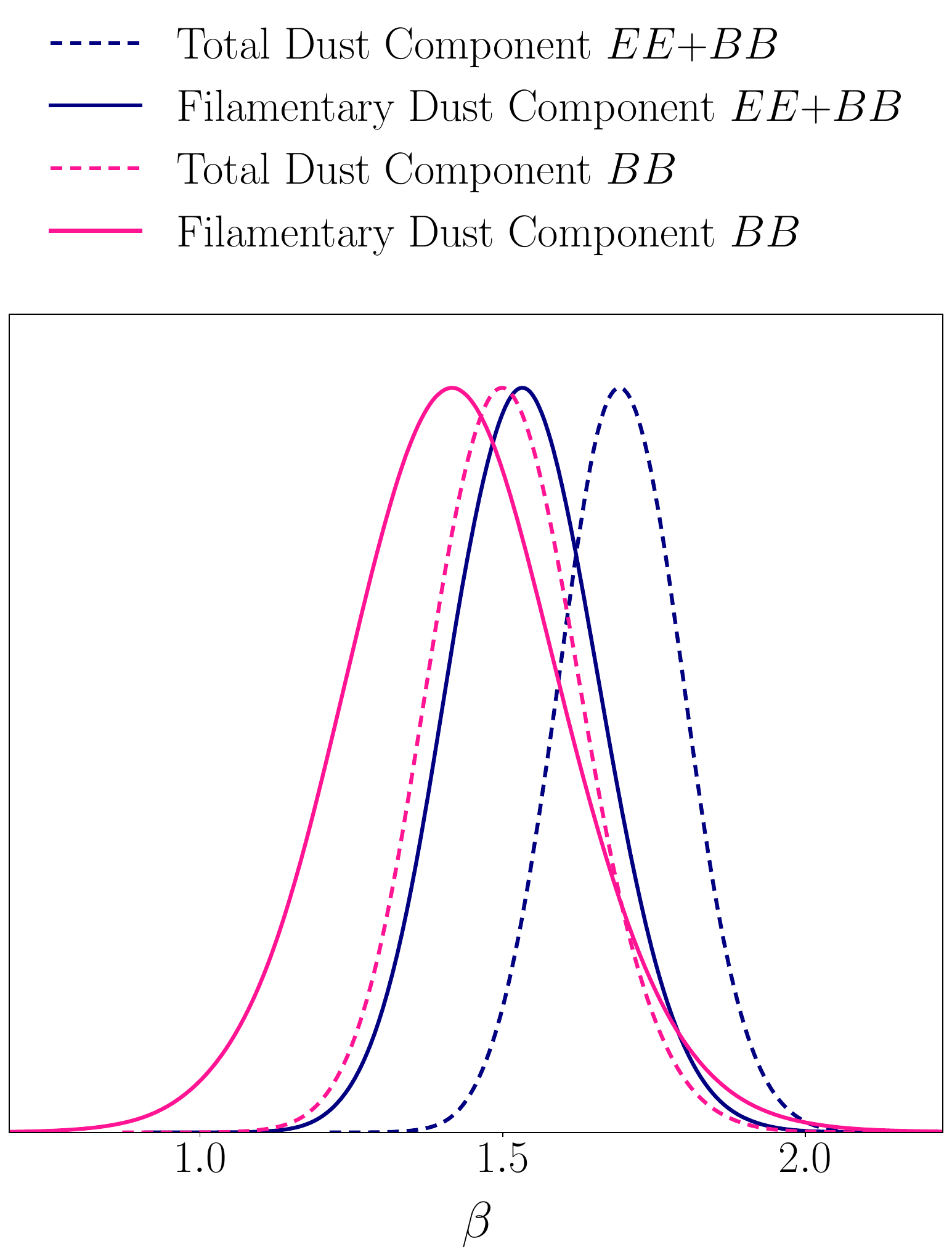} 
\caption{
Comparison of the posteriors for $\beta_{\rm HI}$ through a $\chi^2$ likelihood using cross correlations with the H~\textsc{i} morphology template (solid) to the ones of $\beta_{\rm d}$ using the Hamimeche and Lewis (HL) likelihood with a multicomponent model and no H~\textsc{i} morphology template (dashed). We show the posteriors using $B$~modes only (pink), and $B$~ and $E$~modes (blue). The solid posteriors are the same as in \figref{fig:corner} plotted with the same colors. The $B$-mode-only total dust component posterior is identical to the posterior shown in black in Figure 4 of BK18. \label{fig:beta}}
\end{figure}

From the $\chi^2$ minimization described in \secref{subsubsec:chisqlike}, we get~$\chi^2/$d.o.f. = 1.4, where d.o.f. is the number of degrees of freedom. We find $\hat{\beta}_{\rm HI} = 1.52 \pm 0.11$ and plot the best-fit $BB$ observables for the 4 most sensitive bands used in this measurement in \figref{fig:BBcomps}. The error bars are the square root of the diagonal elements of the covariance matrix used in the $\chi^2$ fit. Since the H~\textsc{i} morphology template does not correlate with the lensed-$\Lambda$CDM, noise, and GD components, the mean of these cross spectra plotted in light green is statistically consistent with zero. Any visible deviations are due to the sample variance in the finite simulation ensemble. The $55<\ell<90$ bandpower of the cross spectrum between the real data and the H~\textsc{i} morphology template fluctuates low relative to the cross spectrum between the H~\textsc{i} morphology template and the H~\textsc{i}-correlated component of the simulation, which is modified by the transfer function defined in \secref{subsec:tf}. This is consistent across frequencies because each multipole bin bandpower is well correlated with the bandpower of the same multipole bin at the different frequencies. The modified H~\textsc{i}-correlated component of the simulation is not guaranteed to match the real data, because we do not have a data-driven model for the multipole-dependent representation of the H~\textsc{i} morphology template in the real dust field (\secref{subsec:tf}). Note that the cross correlations with the real data highly exceed the spurious correlations across all frequencies. 

Taking a Bayesian approach, we use \textsc{cobaya} \citep{2019ascl.soft10019T,2021JCAP...05..057T} to run MCMC and compute the posteriors on $a$, $k$, and $\beta_{\rm HI}$ as described in \secref{subsubsec:chisqlike}. We marginalize over $a$ because the GD cross spectra with the H~\textsc{i} morphology template have no constraining power for $a$, and show the contour plots for the more interesting $k$ and $\beta_{\rm HI}$ in \figref{fig:corner} for $E$~modes only, $B$~modes only, and $E$~and $B$~modes simultaneously. The value for $k$ folds in the normalization of the H~\textsc{i} morphology template. However, the more standard deviations away from zero it is, the stronger the detection of an H~\textsc{i}-correlated component there is in the cross spectra of the real data with the H~\textsc{i} morphology template. The posterior of $\beta_{\rm HI} = 1.54 \pm 0.13$ when using $E$~and $B$~modes simultaneously is consistent with the best-fit value and standard deviation we get using the frequentist maximum-likelihood approach.

We find consistency between the spectral index of the filamentary dust SED, $\beta_{\rm HI}$, and the total dust SED,~$\beta_{\rm d}$, as obtained in BK18 by fitting BICEP/Keck, WMAP and Planck $B$-mode auto and cross spectra to a GD model. That work used a multicomponent parametric model with the Hamimeche and Lewis (HL) likelihood that includes auto and cross spectra across frequencies. The posteriors are shown in \figref{fig:beta} with repeated posteriors from \figref{fig:corner} for comparison. The posteriors plotted are measuring a related but different quantity, because we are correlating with a filament-based template in this paper. The results obtained in BK18 are based on a dust model that assumes a constant ratio between the dust $EE$ and $BB$ power spectra. In this paper, we modify this model to allow the dust $EE$ and $BB$ power spectra to have independent power-law spectral behavior. We find a slight shift to higher values when~$E$~modes are included in the fit. The best-fit values and~$1\sigma$ deviations for the filamentary and total dust components, respectively, are $1.42\pm0.19$ and $1.49\pm0.13$ for $BB$ and $1.54\pm0.13$ and $1.70\pm0.10$ for $EE+BB$. We do not find significant tension between the filamentary and total dust SEDs. However, it would be interesting to check whether the differences become statistically significant with tighter uncertainties, which would have important implications for $B$-mode cosmology.

Since the H~\textsc{i} morphology model is identifying only filamentary contributions to the dust polarization, the similarity in the best-fit values and posteriors for $\beta$ between the two methods indicates that there is no evidence of decorrelation between the filamentary structures that are preferentially associated with the cold neutral medium \citep{Clark:2019,2020A&A...639A..26K} and the rest of the dust column in the BICEP/Keck region. If the H~\textsc{i} morphology method yielded a different SED, the combination of H~\textsc{i} and GD would produce different polarization angles at different frequencies due to the changing relative weighting between the two components. We also find that the results for $\beta_{\rm HI}$ are consistent for different RHT parameters.

The fact that we find a similar SED fit for the filamentary component and for the total dust in the BICEP/Keck region does not have to be the case in other regions of the sky. The dust associated with the warmer, more diffuse H~\textsc{i} component may scale differently in frequency in other regions. Because the H~\textsc{i} morphology templates use the orientation of filamentary structures, a data-driven model for the dust polarization associated with the diffuse, nonfilamentary dust is currently lacking.

\subsection{Individual Frequency Band Contribution \label{subsec:bkvp}}
We study the contribution of each band and instrument used in the results of Sections \ref{subsec:v1} and \ref{subsec:sed} and measure the statistical significance of the detection of filamentary dust polarization as a function of frequency. 

\begin{table}
\centering
\begin{tabular}{c|c|c|c}
 & $BB$ & $EE$ & $BB$ + $EE$ \\
\tableline
BICEP3 95~GHz & 4.53 & 1.22 & 4.72 \\
\tableline
Planck 143~GHz & 0.05 & 0.72 & 0.12 \\
\tableline
BICEP2/Keck 150~GHz & 5.31 & 2.43 & 5.98 \\
\tableline
Planck 217~GHz & 3.50 & 2.37 & 4.02 \\
\tableline
Keck 220~GHz & 5.82 & 7.13 & 9.26 \\
\tableline
Planck 353~GHz & 3.18 & 7.99 & 8.59 \\
\end{tabular}
\caption{Comparison of the statistical significance of a detection of the cross correlation between H~\textsc{i} morphology template and the dust polarization at different frequencies in units of equivalent Gaussian standard deviations as defined in \secref{subsec:statTests}.}
\label{tab:sigPvBK}
\end{table}

We measure a significant detection of dust down to~95~GHz as shown in \tabref{tab:sigPvBK}. These results are insensitive to the covariance matrix conditioning, frequency scaling law, or use of a transfer function for the H~\textsc{i} morphology template as shown in Table \ref{apptab:sigPvBK}. We find that, in the BICEP/Keck region, the BICEP3 95~GHz band is more sensitive to dust polarization than any of the Planck bands below 353~GHz when using both $E$~and $B$~modes and is more sensitive than any Planck band when using $B$~modes only. When using both $E$~and $B$~modes, the Planck 353 GHz band is the only Planck band that exceeds 5$\sigma$, while the 150 and 220~GHz bands of BICEP/Keck both exceed 5$\sigma$, and the 95~GHz band is correlated with the H~\textsc{i} morphology template at $\sim$5$\sigma$. This shows the power of the BICEP/Keck bands for characterizing the dust in this field, and especially, for measuring its SED. The detection at 95~GHz is also interesting because it provides a low-frequency lever arm for the dust SED, and it is the band where the $\Lambda$CDM component starts to dominate over the dust component in polarized emission at smaller scales \citep{2021PhRvL.127o1301A}. These results are consistent with our expectations from the map depths we have shown in BK18 and with the number of standard deviations away from zero the peak of the posterior for $k$ is for each case.

At frequencies lower than~$220$~GHz, almost all of the detection significance is coming from $B$~modes. That is, the statistical significance of the detection is equivalent at lower frequencies when including $E$~modes. This is because at lower frequencies, in $E$~modes, we are limited by the sample variance of the CMB, i.e., the statistical significance of the detection will not improve unless we remove the CMB component or increase the observed sky area. The $E$~modes at those frequencies produce a negligible change in the overall significance estimates because they are downweighted by our statistical metrics.

\begin{figure}[t!]
\plotone{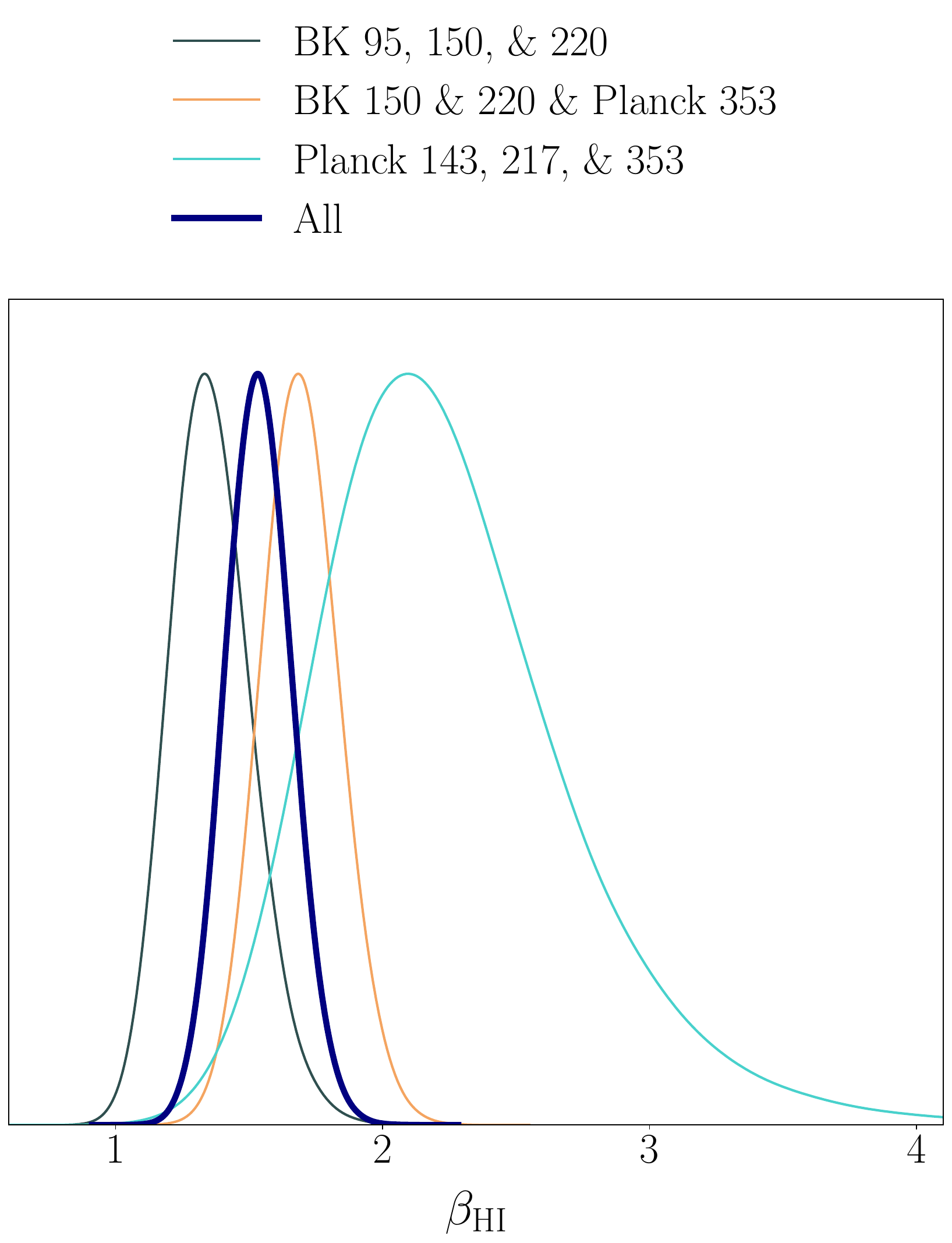}
\caption{
Comparison of the posteriors for $\beta_{\rm HI}$ we get through a $\chi^2$ likelihood using $E$- and $B$-mode cross correlations with the H~\textsc{i} morphology template for different selections of frequency bands and for BICEP/Keck only and Planck only variations. The thick navy posterior labeled ``All" is the same as the navy posterior in Figures \ref{fig:corner} and \ref{fig:beta}.
\label{fig:beta_lovhi}}
\end{figure}

Moreover, we can measure whether the SED changes when we omit the low- or high-frequency channels from our analysis. We show the $\beta_{\rm HI}$ posteriors, using both $E$~and $B$~modes in the fits, in \figref{fig:beta_lovhi}. For the BICEP/Keck-only case, we find~$\beta_{\rm HI} = 1.36^{+0.14}_{-0.17}$. For the Planck-only case, we find~$\beta_{\rm HI} = 2.26^{+0.32}_{-0.54}$. Using similar frequencies to the Planck-only case but replacing Planck's 143 and 217~GHz bands with the 150 and 220~GHz bands of BICEP/Keck, we find~$\beta_{\rm HI}~=~1.69~\pm~0.15$. Finally, we also plot the posterior using all the frequency bands, for which~$\beta_{\rm HI}~=~1.54~\pm~0.13$, with the same color as in Figures \ref{fig:corner} and \ref{fig:beta} for comparison.

Note that, although two of the cases cover approximately the same frequency range, the Planck-only case has a wider posterior that is shifted slightly toward higher values of $\beta_{\rm HI}$. This is because Planck's 143 and 217~GHz bands are not very sensitive to filamentary dust polarization when restricted to the BICEP/Keck region as compared to BICEP/Keck's 150 and 220~GHz bands. That said, the four posteriors are statistically consistent with each other to within~$2\sigma$. The results are qualitatively similar when fitting $E$~modes and $B$~modes separately. 

Finally, we also calculate the correlation ratio as a function of multipole $\ell$ between BICEP/Keck or Planck data and V1 with RHT parameters from \eqref{eq:bestparams}. The correlation ratio is defined as
\eq{
\rho_\ell^{{\rm data} \times {\rm HI}} = \frac{D_\ell^{{\rm data} \times {\rm HI}}}{
\sqrt{D_\ell^{{\rm data} \times {\rm data}} \times D_\ell^{{\rm HI} \times {\rm HI}}}}. \label{eq:corr_rat}
}
The autospectra in the denominator contain noise biases. It would be possible to debias, but this would change the interpretation of the resulting correlation ratio. With noise debiasing, the correlation ratio would reflect the fraction of the sky signal that is accounted for by the H~\textsc{i} morphology template. Without noise debiasing, as in \eqref{eq:corr_rat}, the correlation ratio reflects the fraction of the data (including noise) that is accounted for by the H~\textsc{i} morphology template. For the purposes of forecasting sensitivity to $r$, we wish to retain the diluting effects of noise.

\begin{figure*}[t!]
\plotone{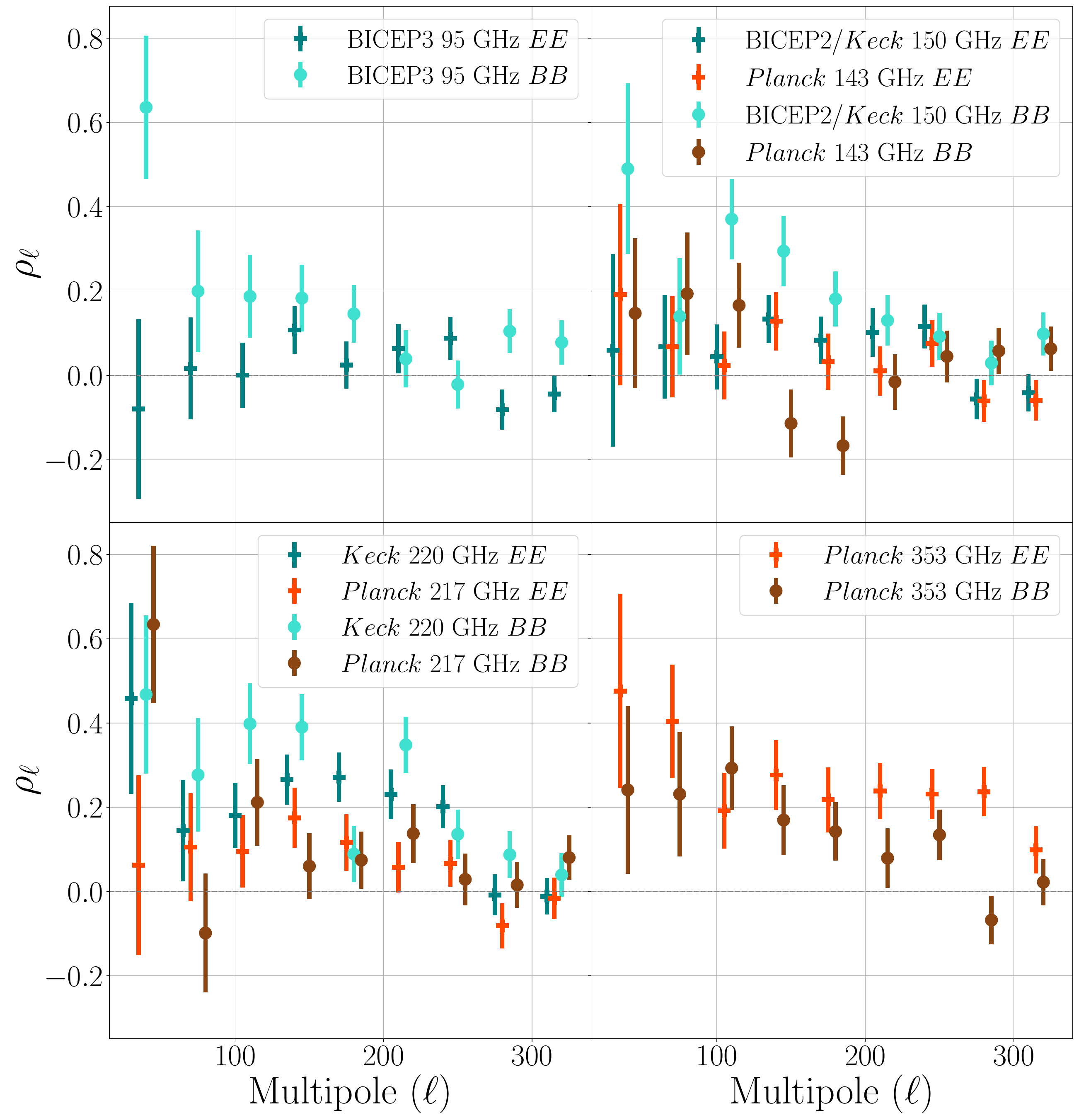}
\caption{
$EE$ (cross) and $BB$ (circle) unitless correlation ratios as a function of multipole moment. The correlation ratios between V1 and Planck data with 1$\sigma$ variations are shown in red and brown to compare them to the correlation ratios between V1 and BICEP/Keck data, which are shown in teal and turquoise. The errors are derived from spurious correlations between V1 and lensed-$\Lambda$CDM, Gaussian dust, and noise. Data points for similar frequencies between BICEP/Keck and Planck are plotted on the same panels for comparison.
\label{fig:rells}}
\end{figure*}

\begin{figure*}[t!]
\plotone{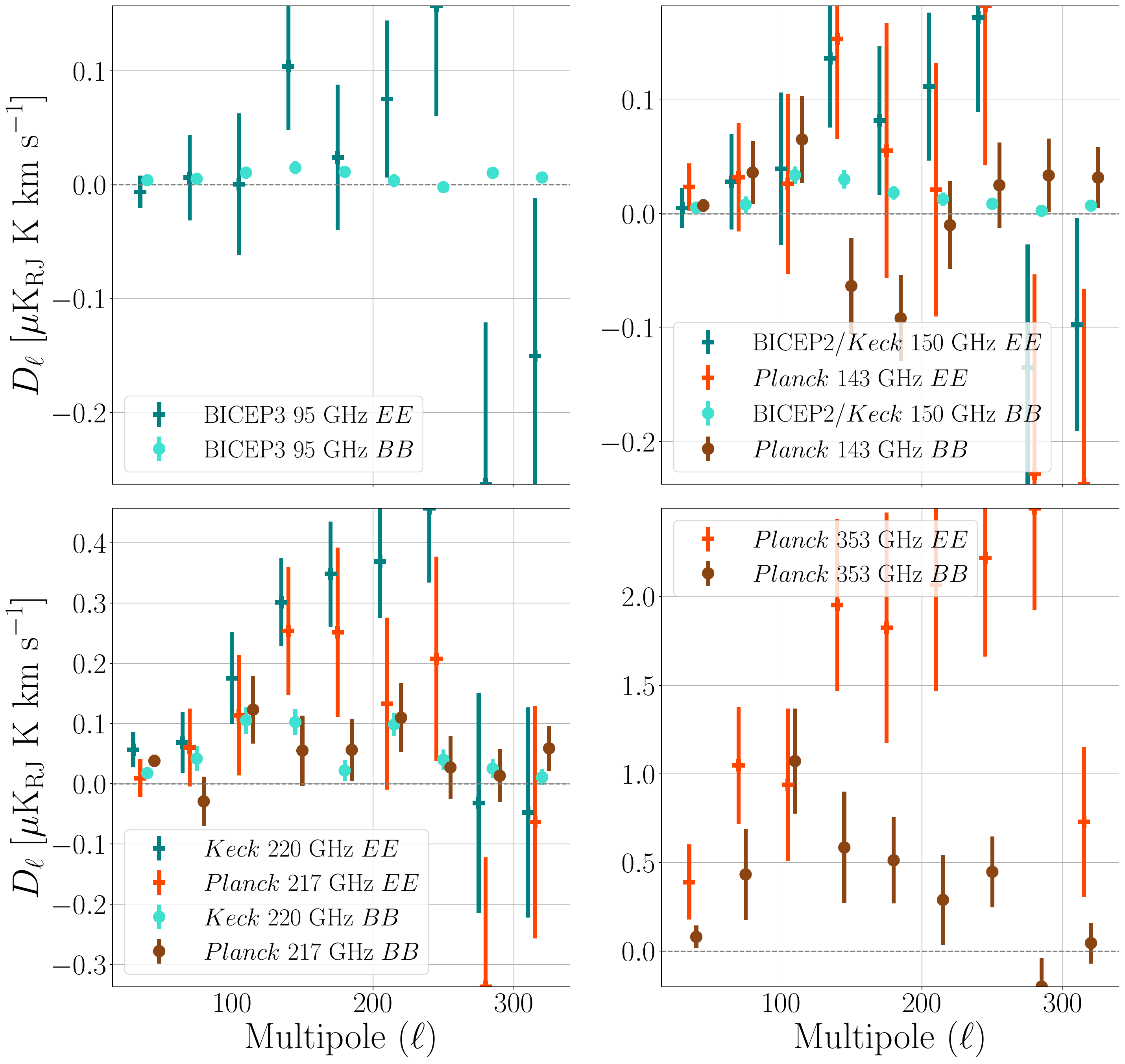}
\caption{
$EE$ (cross) and $BB$ (circle) cross spectra as a function of multipole moment. The cross spectra between V1 and Planck data with 1$\sigma$ variations are shown in red and brown to compare them to the cross spectra between V1 and BICEP/Keck data, which are shown in teal and turquoise. The errors are derived from spurious correlations between V1 and lensed-$\Lambda$CDM, Gaussian dust, and noise. Data points for similar frequencies between BICEP/Keck and Planck are plotted on the same panels for comparison.
\label{fig:Dells}}
\end{figure*}

We plot the results in \figref{fig:rells}. The error bars show the 1$\sigma$ deviation of the correlation of 499 realizations of lensed-$\Lambda$CDM, GD, and noise with V1. Comparing BICEP/Keck data points with Planck bands of similar frequencies, we note that the BICEP/Keck bands correlate better in $B$ modes with the H~\textsc{i} morphology template in this region. Also, the BICEP/Keck 220~GHz data is only slightly less correlated with V1 in $EE$ but much more correlated in $BB$ than the Planck 353~GHz data. This is consistent with the dust sensitivity estimates from BK18 that show that the BICEP/Keck 220~GHz data is more sensitive to dust than the Planck 353~GHz data (Figure 6 of BK18). The correlation ratio is larger in $B$ modes than that in $E$ modes for BICEP/Keck bands due to the CMB sample variance at lower frequencies. For a direct comparison of the error bars between BICEP/Keck and Planck bands of similar frequencies, we plot the numerator of the correlation ratio $\rho_\ell$, i.e., the cross spectra $D_\ell$ in \figref{fig:Dells}. The error bars are clearly smaller for the BICEP/Keck bands, especially in $B$ modes.

\subsection{Polarized Dust in Magellanic Stream~\textsc{i} \label{subsec:magstream}}
Passing through the BICEP/Keck region is a stream of high-velocity gas, known as Magellanic Stream~\textsc{i} \citep{2018MNRAS.474..289W}. The metallicity and abundance measurements of the Magellanic Stream are consistent with an origin in the SMC, created by a gravitational tug from the Large Magellanic Cloud \citep{Fox_2018}. The Magellanic Stream and Clouds are part of the Magellanic System, along with the Magellanic Bridge and the Leading Arm \citep[see, e.g.,][for a review]{2016ARA&A..54..363D}. The nature of dust in the Magellanic Stream is not well constrained by observations. Measurements of the gas-to-dust ratio in the Magellanic Clouds indicate a much lower dust content than that in the Milky Way \citep{fong1987deep,richter2000orfeus,Tumlinson_2002}. 

However, there is good reason to believe that the Stream may contain some dust since the same processes that inject metals, such as Mg~\textsc{ii} and Fe~\textsc{ii}, into clouds should also inject dust \citep{Benjamin2005,2001}. Constraints on the dust content of the Magellanic Stream can thus have important implications for dust survival and destruction in the Stream environment. Although efforts to detect dust emission from the Magellanic Stream have not yielded positive results yet in intensity or reddening \citep{1986A&A...170...84W,2017ApJ...846...38L}, we test whether we can detect it in polarization, assuming the dust there is polarized due to a coherent magnetic field. While not yet directly detected in the Magellanic Stream, a coherent magnetic field is plausible given the detections in other tidal features and in the Magellanic Bridge using Faraday rotation measurements \citep{2017MNRAS.467.1776K}.

\begin{table}
\centering
\begin{tabular}{c|c|c|c|c}
 & V1 & V2 & V3 & V2 + V3 \\
\tableline
$BB$ & 6.7 & 1.3 & 0.6 & 0.9 \\
\tableline
$EE$ & 14.6 & 2.4 & 1.4 & 2.5 \\
\tableline
$BB + EE$ & 16.1 & 1.6 & 1.2 & 0.3 \\
\end{tabular}
\caption{Comparison of the statistical significance of a detection of the cross correlation between H~\textsc{i} morphology templates and the dust polarization in units of equivalent Gaussian standard deviations for V1, V2, and V3. We also add a column for V2 + V3, both of which are associated with Magellanic Stream~\textsc{i}. The 95, 150, and 220 GHz bands of BICEP/Keck and the 353 GHz band of Planck are used here.}
\label{tab:sigV123}
\end{table}

Using the $\Delta\chi^2$ statistical test defined in \secref{subsubsec:DeltaChi2}, we find no statistically significant correlation with the second and third velocity components, V2 and V3. The correlation metric does not exceed $\sim$2.5$\sigma$ for any of the choices in \tabref{tab:sigV123}, including the addition of V2 and V3. This is also true for all the different variations of RHT parameters we tried. 

We also try looking for a correlation in total intensity ($TT$) between BICEP/Keck or Planck $T$ and V2 or V3 $T$ (the H~\textsc{i} intensity integrated over the V2 and V3 velocity ranges) and find no correlation. Since the standard BICEP/Keck simulations constrain $T$ to the well-measured Planck $T$ map, we only use that one realization for computational simplicity and only look for a visual correlation rather than making statistical inferences. Furthermore, adding V2, V3, or both to V1 decreases the $TT$ correlation with BICEP/Keck and Planck.

We therefore do not detect evidence for dust in Magellanic Stream~\textsc{i}. The Magellanic Stream's distance may limit our sensitivity to resolving the local magnetic field orientations there because structures on the plane of the sky of the same angular scale as the Galactic gas correspond to much larger structures at the distance of the Magellanic Stream. The Stream's distance is fairly uncertain. \citet{2021ApJ...921L..36L} recently estimated it to be $\sim$20~kpc away from the Sun at its closest point through the use of simulations. For comparison, at these high-Galactic latitudes, the dust associated with V1 is likely at a distance of order 100~pc \citep[e.g.,][]{2022A&A...664A.174V,2021ApJ...906...47G}. Furthermore, our analysis is restricted to only the section of the Stream that intersects the BICEP/Keck region. Extending the sky area to include the entire Stream, running the RHT on forthcoming H~\textsc{i} emission data from the Galactic Australian Square Kilometre Array Pathfinder (GASKAP) Survey \citep{GASKAP} with $30''$ angular resolution, and using higher angular resolution dust polarization data \citep{2021arXiv210710364C,2022ApJ...929..166H,Abazajian_2022} are all possible extensions of this work that can improve the sensitivity of this method for detecting or setting limits on dust polarization from the Stream.

\section{Summary and Outlook} \label{sec:conc}
We characterize the filamentary dust polarization in the BICEP/Keck observing region through correlations with template maps based on measurements of H~\textsc{i}. A detection of primordial gravitational wave $B$~modes depends on reliable component separation because the polarized dust emission is the dominant foreground at frequencies~$\gtrapprox$ 70~GHz \citep{2009AIPC.1141..222D,2016A&A...594A..10P} and has a higher amplitude than that of the polarization associated with primordial gravitational waves \citep{2014JCAP...08..039F,2015PhRvL.114j1301B,2016JCAP...03..052E}. Therefore, polarized dust emission must be characterized to great accuracy and precision. We concentrate on the BICEP/Keck region as a test case for the diffuse high-Galactic latitude sky with deep data across several frequencies.

We summarize the conclusions of this work below.
\begin{itemize}
\item We separate the H~\textsc{i} emission in the BICEP/Keck region into three distinct velocity components that together account for the bulk of the polarized intensity in the H~\textsc{i} morphology template. One is associated with the Milky Way, while the other two are associated with Magellanic Stream~\textsc{i}.
\item We explore the RHT parameter space to increase the correlation with BICEP/Keck and Planck by $\sim$2$\sigma$ in $BB$ and $\sim$3$\sigma$ in $EE$ and $EE$+$BB$ with respect to the parameters used in \citet{Clark:2019}. The parameters we recommend using on H\textsc{i}4PI data in the BICEP/Keck region for producing H~\textsc{i} morphology templates are $D_W = 135'$, $\theta_{\rm FWHM} = 4'$, and $Z = 0.75$.
\item Using polarization data from BICEP/Keck and Planck, we find a statistically significant detection of filamentary dust polarization in the Galactic component of H~\textsc{i} at $\sim$7$\sigma$ in $BB$, $\sim$15$\sigma$ in $EE$, and $\sim$16$\sigma$ in $EE$+$BB$.
\item We show that the overwhelming majority of the contribution comes from the LVC velocity range, $-12~{\rm km~s}^{-1}~<~v_{\rm lsr}~<~10$~km~s$^{-1}$, and find no evidence of frequency decorrelation in the BICEP/Keck region as defined in \citet{2021A&A...647A..16P}. The inclusion of the IVC component to the line-of-sight sum affects the correlation by $\lesssim 0.1 \sigma$ in $BB$, $\lesssim 0.3 \sigma$ in $EE$, and $\lesssim 0.2 \sigma$ in $EE + BB$. We note that the dust structure associated with H~\textsc{i} kinematic substructure within the LVC range could still produce frequency decorrelation.
\item We fit an SED with $\hat{\beta}_{\rm HI} = 1.54 \pm 0.13$ in the BICEP/Keck region for the filamentary dust polarization component associated with the Galactic component. This is consistent with the SED fit in BK18 for the total dust component in the BICEP/Keck region. The similarity between the SED of the filamentary contributions to the dust polarization and the SED of the rest of the dust field indicates that there is no evidence for decorrelation between the filamentary dust and the rest of the dust column in the BICEP/Keck region.

\item We present the first multifrequency detection of filamentary dust polarization in cross-correlation with H~\textsc{i} filaments down to 95~GHz. We show that the 95~GHz band of BICEP3 is more sensitive than any Planck band to the $B$-mode correlation in the BICEP/Keck region, providing a low-frequency lever arm for the dust SED. We also find that, at low frequencies, the brightness of the CMB in $E$~modes limits our sensitivity but that the correlation could improve in $B$~modes with more data. As a consistency check, we also omit certain frequency bands in the multifrequency correlations to compare the contribution of the different bands to our measurements.

\item We do not find evidence for dust polarization in the higher-velocity H~\textsc{i} components associated with Magellanic Stream~\textsc{i}. This confirmation is important for future CMB observations whose field-of-view intercepts the Magellanic Stream. 
\end{itemize}

In addition to facilitating foreground removal for $B$-mode cosmology, this type of H~\textsc{i}-based characterization of the dust polarization can also be a method for removing the Milky Way foreground contribution for studies of the Magellanic Clouds in dust polarization. Such a study is planned with CCAT-prime \citep{2021arXiv210710364C}.
\\\\

\begin{acknowledgments}
We thank the anonymous referee for a thoughtful review.

This work was supported by the National Science Foundation under grant No. AST-2106607.

The BICEP/Keck projects have been made possible through a series of grants from the National Science Foundation including 0742818, 0742592, 1044978, 1110087, 1145172, 1145143, 1145248, 1639040, 1638957, 1638978, and 1638970, and by the Keck Foundation. We thank the staff of the U.S. Antarctic Program and in particular the South Pole Station and the heroic winter-overs without whose help this research would not have been possible. We also thank all those who have contributed past efforts to the BICEP/Keck series of experiments. 

This publication utilizes data from Planck, an ESA science mission funded by ESA Member States, NASA, and Canada.

This work makes use of data from the H\textsc{i}4PI Survey, which is constructed from the Effelsberg-Bonn H~\textsc{i} Survey (EBHIS), made with the 100 m radio telescope of the MPIfR at Effelsberg/Germany, and the Galactic All-Sky Survey (GASS), observed with the Parkes Radio Telescope, part of the Australia Telescope National Facility, which is funded by the Australian Government for operation as a National Facility managed by CSIRO. EBHIS was funded by the Deutsche Forschungsgemein-schaft (DFG) under the grants KE757/7-1 to 7-3.

The computations in this paper were run on the Sherlock cluster, supported by the Stanford Research Computing Center at Stanford University, and on the Odyssey/Cannon cluster, supported by the FAS Science Division Research Computing Group at Harvard University.
\end{acknowledgments}

\vspace{5mm}

\software{astropy \citep{2013A&A...558A..33A,2018AJ....156..123A},
          Healpix\footnote{\url{http://healpix.sourceforge.net/}} \citep{2005ApJ...622..759G},
          healpy \citep{2019JOSS....4.1298Z},
          matplotlib \citep{2007CSE.....9...90H},
          numpy \citep{10.5555/2886196},
          \textsc{cobaya} \citep{2019ascl.soft10019T,2021JCAP...05..057T},
          GetDist \citep{2019arXiv191013970L}
          }

\appendix

\section{Uncertainty Calculation} \label{app:unc}
To measure the uncertainty on the best-fit $a$, $k$, and $\beta_{\rm HI}$ values, we construct a simulation set of 499 filtered dust realization Stokes $Q$/$U$ maps as
\eq{ \tilde{m}_\nu\p{\mathrm{dust}}(\unitVec{n}, \alpha) \equiv \alpha \cdot f_\nu(\beta_{\rm GD}) \cdot \tilde{m}_\nu\p{\mathrm{GD}} (\unitVec{n}) + \hat{k} \cdot f_\nu(\hat{\beta}_{\rm HI}) \cdot \tilde{m}_\nu\p{\tilde{\mathrm{HI}}}(\unitVec{n}) , }
where $f_\nu$ is a modified blackbody scaling law with a fixed temperature, $T =  19.6\,{\rm K}$, as in \secref{subsec:sims}, $\hat{k}$ and $\hat{\beta}_{\rm HI}$ are the best-fit results from the real data, and $\tilde{m}_\nu\p{\tilde{\mathrm{HI}}}$ is the result of applying the transfer function defined in \secref{subsec:tf} in harmonic space to $\tilde{m}_\nu\p{\mathrm{HI}}$ and then inverse transforming back to map space. The free parameter $\alpha$ is chosen such that
\eq{
\overline{\mathbf{D}}^{{\rm dust}\times{\rm dust}}(\alpha) 
= f^2_\nu(\beta_{\rm GD}) \overline{\mathbf{D}}^{{\rm GD}\times{\rm GD}},
\label{eq:cond2}
}
where $\overline{\mathbf{D}}$ is the mean over realizations of the vector of autospectra over $EE$, $BB$, and multipole bins. One frequency, 353~GHz, is sufficient for the fit here.

Therefore, we fit for $\alpha$ using a Gaussian likelihood approximation, i.e. a $\chi^2$-minimization
\eq{-2 \log \mathcal{L} = \parens{ \hat{\mathbf{S}}(\alpha) - \mathbf{S} }^T \mathbf{Z}^{-1} \parens{ \hat{\mathbf{S}}(\alpha) - \mathbf{S} },}
where, from Equation \ref{eq:cond2},
\begin{eqnarray}
\hat{\mathbf{S}}(\alpha) - \mathbf{S} =&& (\alpha^2 f^2_{{\rm 353}\,{\rm GHz}}(\beta_{\rm GD})  - 1) \cdot \overline{\mathbf{D}}^{{\rm GD}\times{\rm GD}} + \\
&&+ \hat{k}^2 \cdot f^2_{{\rm 353}\,{\rm GHz}}(\hat{\beta}_{\rm HI}) \cdot \mathbf{D}^{\tilde{{\rm HI}}\times{\tilde{\rm HI}}} + \nonumber\\
&&+ 2 \cdot \alpha \cdot \hat{k} \cdot f_{{\rm 353}\,{\rm GHz}}(\hat{\beta}_{\rm HI}) \cdot f_{{\rm 353}\,{\rm GHz}}(\beta_{\rm GD}) \cdot \overline{\mathbf{D}}^{{\rm GD}\times{\tilde{\rm HI}}} \nonumber
\end{eqnarray}
and $\mathbf{Z}$ is the covariance matrix due to variations in the GD. 

After fitting $\alpha$, we define
\eq{d_\nu =  \tilde{m}\p{\mathrm{\Lambda CDM}}_\nu(\unitVec{n}) + \tilde{m}_\nu\p{n}(\unitVec{n}) + \tilde{m}\p{\mathrm{dust}}_\nu(\unitVec{n},\hat{\alpha}),}
where $\hat{\alpha}$ is the best-fit value, and repeat the process in \secref{subsubsec:chisqlike}, replacing $\mathbf{D}\p{\rm real}$ with the cross spectra of $d_\nu$ with the H~\textsc{i} morphology template.

Expecting the fits for $\alpha$, $k$, and $\beta_{\rm HI}$ to yield the inputs $\hat{\alpha}$, $\hat{k}$, and  $\hat{\beta}_{\rm HI}$, we use the spread of the best-fit distributions for the 499 realizations to calculate the uncertainty on our fitting method for $a$, $k$, and $\beta_{\rm HI}$, respectively. An example of this is shown in \figref{fig:unc} and described in \secref{subsec:uncertainty}.

\section{Analysis Variations} \label{app:variations}
For the main results presented in Tables~\ref{tab:sigdefbest}, \ref{tab:sigbound}, \ref{tab:sigPvBK}, and \ref{tab:sigV123}, we condition the covariance matrix and use a transfer function and a modified blackbody scaling for the H~\textsc{i} morphology template. In this appendix, we present those same results for different variations of those choices in Tables~\ref{apptab:sigdefbest}, \ref{apptab:sigbound}, \ref{apptab:sigPvBK}, \ref{apptab:sigV123}, respectively. The main results are shown in the bolded columns of these tables. Note that the results are not qualitatively affected by these variations.

\begin{deluxetable*}{cc|cccccccc}
\tablecaption{Statistical significance of the detection of V1 in units of equivalent Gaussian standard deviations as defined in \secref{subsubsec:DeltaChi2} using the 95, 150, and 220 GHz bands of BICEP/Keck and the 353 GHz band of Planck. The rows labeled ``best" use the parameters $D_W = 135'$, $\theta_{\rm FWHM} = 4'$, and $Z = 0.75$, and the rows labeled ``default" use the parameters $D_W = 75'$, $\theta_{\rm FWHM} = 30'$, and $Z = 0.7$, which are used in \citet{2019ApJ...887..136C}. The bolded column (9) shows the main results. The other columns show the results for different variations of our model.
\label{apptab:sigdefbest}}
\tablewidth{0pt}
\tablehead{
\multicolumn2c{Covariance Matrix:} & \multicolumn4c{Not Conditioned} & \multicolumn4c{Conditioned}\\
\multicolumn2c{Frequency Scaling:} & \multicolumn2c{Power Law} & \multicolumn2c{Modified Blackbody} & \multicolumn2c{Power Law} & \multicolumn2c{Modified Blackbody}\\
\multicolumn2c{Transfer Function:} & \colhead{Used} & \colhead{Not Used} & \colhead{Used} & \colhead{Not Used} & \colhead{Used} & \colhead{Not Used} & \colhead{Used} & \colhead{Not Used}
}
\decimalcolnumbers
\startdata
$BB$ & best & 7.0 & 6.3 & 6.8 & 6.2 & 6.9 & 6.2 & {\bf 6.7} & 6.2 \\
 & default & 4.8 & 3.9 & 4.7 & 3.9 & 4.7 & 4.0 & {\bf 4.7} & 4.0 \\
\hline
$EE$ & best & 15.2 & 14.9 & 15.2 & 14.9 & 14.6 & 14.2 & {\bf 14.6} & 14.2 \\
 & default & 12.2 & 10.7 & 12.2 & 10.7 & 12.3 & 10.8 & {\bf 12.3} & 10.8 \\
\hline
$BB + EE$ & best & 17.2 & 16.6 & 17.1 & 16.5 & 16.2 & 15.7 & {\bf 16.1} & 15.6 \\
 & default & 13.8 & 12.3 & 13.8 & 12.3 & 12.9 & 11.6 & {\bf 12.9} & 11.6 \\
\enddata
\end{deluxetable*}

\begin{deluxetable*}{cc|cccccccc}
\tablecaption{
Comparison of the statistical significance of a detection of the cross correlation with the dust polarization in units of equivalent Gaussian standard deviations when including the channels in the IVC velocity range in the line-of-sight sum. The RHT parameters from \eqref{eq:bestparams} are used here for the H~\textsc{i} morphology template with the 95, 150, and 220 GHz bands of BICEP/Keck and the 353 GHz band of Planck. The bolded column (9) shows the main results. The other columns show the results for different variations of our model.
\label{apptab:sigbound}}
\tablewidth{0pt}
\tablehead{
\multicolumn2c{Covariance Matrix:} & \multicolumn4c{Not Conditioned} & \multicolumn4c{Conditioned}\\
\multicolumn2c{Frequency Scaling:} & \multicolumn2c{Power Law} & \multicolumn2c{Modified Blackbody} & \multicolumn2c{Power Law} & \multicolumn2c{Modified Blackbody}\\
\multicolumn2c{Transfer Function:} & \colhead{Used} & \colhead{Not Used} & \colhead{Used} & \colhead{Not Used} & \colhead{Used} & \colhead{Not Used} & \colhead{Used} & \colhead{Not Used}
}
\decimalcolnumbers
\startdata
$BB$ & range for LVCs + IVCs & 7.0 & 6.3 & 6.8 & 6.2 & 6.9 & 6.2 & {\bf 6.7} & 6.2 \\
 & range for LVCs & 7.1 & 6.4 & 7.0 & 6.3 & 7.0 & 6.2 & {\bf 6.8} & 6.2 \\
\hline
$EE$ & range for LVCs + IVCs & 15.2 & 14.9 & 15.2 & 14.9 & 14.6 & 14.2 & {\bf 14.6} & 14.2 \\
 & range for LVCs & 14.9 & 14.8 & 14.9 & 14.7 & 14.4 & 14.2 & {\bf 14.3} & 14.2 \\
\hline
$BB + EE$ & range for LVCs + IVCs & 17.2 & 16.6 & 17.1 & 16.5 & 16.2 & 15.7 & {\bf 16.1} & 15.6 \\
 & range for LVCs & 16.9 & 16.4 & 16.9 & 16.3 & 16.2 & 15.7 & {\bf 16.1} & 15.6 \\
\enddata
\end{deluxetable*}

\begin{deluxetable*}{cc|cccccccc}
\tablecaption{Comparison of the statistical significance of a detection of the cross correlation between H~\textsc{i} morphology templates and the dust polarization at different frequencies in units of equivalent Gaussian standard deviations as defined in \secref{subsec:statTests}. The bolded column (9) shows the main results. The other columns show the results for different variations of our model.
\label{apptab:sigPvBK}}
\tablewidth{0pt}
\tablehead{
\multicolumn2c{Covariance Matrix:} & \multicolumn4c{Not Conditioned} & \multicolumn4c{Conditioned}\\
\multicolumn2c{Frequency Scaling:} & \multicolumn2c{Power Law} & \multicolumn2c{Modified Blackbody} & \multicolumn2c{Power Law} & \multicolumn2c{Modified Blackbody}\\
\multicolumn2c{Transfer Function:} & \colhead{Used} & \colhead{Not Used} & \colhead{Used} & \colhead{Not Used} & \colhead{Used} & \colhead{Not Used} & \colhead{Used} & \colhead{Not Used}
}
\decimalcolnumbers
\startdata
 & BICEP3 95~GHz & 4.44 & 3.84 & 4.44 & 3.84 & 4.53 & 3.98 & {\bf 4.53} & 3.98 \\
 & Planck 143~GHz & 0.16 & 0.36 & 0.16 & 0.36 & 0.05 & 0.40 & {\bf 0.05} & 0.40  \\
$BB$ & BICEP2/Keck 150~GHz & 5.13 & 4.76 & 5.13 & 4.76 & 5.31 & 4.97 & {\bf 5.31} & 4.97  \\
 & Planck 217~GHz & 3.50 & 3.31 & 3.50 & 3.31 & 3.50 & 3.14 & {\bf 3.50} & 3.14  \\
 & Keck 220~GHz & 5.90 & 5.65 & 5.90 & 5.65 & 5.82 & 5.60 & {\bf 5.82} & 5.60  \\
 & Planck 353~GHz & 3.18 & 2.53 & 3.18 & 2.53 & 3.18 & 2.60 & {\bf 3.18} & 2.60  \\
\hline
 & BICEP3 95~GHz & 1.21 & 1.18 & 1.21 & 1.18 & 1.22 & 1.25 & {\bf 1.22} & 1.25 \\
 & Planck 143~GHz & 0.76 & 0.20 & 0.76 & 0.20 & 0.72 & 0.20 & {\bf 0.72} & 0.20 \\
$EE$ & BICEP2/Keck 150~GHz & 2.42 & 2.00 & 2.42 & 2.00 & 2.43 & 2.01 & {\bf 2.43} & 2.01 \\
 & Planck 217~GHz & 2.28 & 1.52 & 2.28 & 1.52 & 2.37 & 1.62 & {\bf 2.37} & 1.62 \\
 & Keck 220~GHz & 7.35 & 6.79 & 7.35 & 6.79 & 7.13 & 6.61 & {\bf 7.13} & 6.61 \\
 & Planck 353~GHz & 7.92 & 8.13 & 7.92 & 8.13 & 7.99 & 8.12 & {\bf 7.99} & 8.12 \\
\hline
 & BICEP3 95~GHz & 4.61 & 3.93 & 4.61 & 3.93 & 4.72 & 4.05 & {\bf 4.72} & 4.05 \\
 & Planck 143~GHz & 0.13 & 1.32 & 0.13 & 1.32 & 0.12 & 1.55 & {\bf 0.12} & 1.55 \\
$BB+EE$ & BICEP2/Keck 150~GHz & 5.83 & 5.32 & 5.83 & 5.32 & 5.98 & 5.49 & {\bf 5.98} & 5.49 \\
 & Planck 217~GHz & 3.72 & 2.80 & 3.72 & 2.80 & 4.02 & 3.05 & {\bf 4.02} & 3.05 \\
 & Keck 220~GHz & 9.02 & 8.46 & 9.02 & 8.46 & 9.26 & 8.76 & {\bf 9.26} & 8.76 \\
 & Planck 353~GHz & 8.65 & 8.59 & 8.65 & 8.59 & 8.59 & 8.57 & {\bf 8.59} & 8.57 \\
\enddata
\end{deluxetable*}

\begin{deluxetable*}{cc|cccccccc}
\tablecaption{
Comparison of the statistical significance of a detection of the cross correlation between H~\textsc{i} morphology templates and the dust polarization in units of equivalent Gaussian standard deviations for V1, V2, and V3. We also add a column for V2 + V3, both of which are associated with Magellanic Stream~\textsc{i}. The 95, 150, and 220 GHz bands of BICEP/Keck and the 353 GHz band of Planck are used here. The bolded column (9) shows the main results. The other columns show the results for different variations of our model.
\label{apptab:sigV123}}
\tablewidth{0pt}
\tablehead{
\multicolumn2c{Covariance Matrix:} & \multicolumn4c{Not Conditioned} & \multicolumn4c{Conditioned}\\
\multicolumn2c{Frequency Scaling:} & \multicolumn2c{Power Law} & \multicolumn2c{Modified Blackbody} & \multicolumn2c{Power Law} & \multicolumn2c{Modified Blackbody}\\
\multicolumn2c{Transfer Function:} & \colhead{Used} & \colhead{Not Used} & \colhead{Used} & \colhead{Not Used} & \colhead{Used} & \colhead{Not Used} & \colhead{Used} & \colhead{Not Used}
}
\decimalcolnumbers
\startdata
 & V1 & 7.0 & 6.3 & 6.8 & 6.2 & 6.9 & 6.2 & {\bf 6.7} & 6.2 \\
$BB$ & V2 & 1.0 & 1.0 & 1.1 & 1.1 & 1.1 & 1.3 & {\bf 1.3} & 1.4 \\
 & V3 & 0.8 & 0.2 & 0.8 & 0.2 & 0.7 & 0.1 & {\bf 0.6} & 0.1 \\
 & V2 + V3 & 0.7 & 0.5 & 0.8 & 0.5 & 0.9 & 0.7 & {\bf 0.9} & 0.7 \\
\hline
 & V1 & 15.2 & 14.9 & 15.2 & 14.9 & 14.6 & 14.2 & {\bf 14.6} & 14.2 \\
$EE$ & V2 & 2.4 & 2.3 & 2.4 & 2.4 & 2.3 & 2.3 & {\bf 2.4} & 2.3 \\
 & V3 & 1.4 & 1.1 & 1.4 & 0.8 & 1.5 & 0.8 & {\bf 1.4} & 1.1 \\
 & V2 + V3 & 1.6 & 0.5 & 1.6 & 0.2 & 2.5 & 2.3 & {\bf 2.5} & 2.3 \\
\hline
 & V1 & 17.2 & 16.6 & 17.1 & 16.5 & 16.2 & 15.7 & {\bf 16.1} & 15.6 \\
$BB + EE$ & V2 & 2.0 & 2.2 & 2.1 & 2.3 & 1.5 & 2.1 & {\bf 1.6} & 2.1 \\
 & V3 & 1.1 & 1.1 & 1.2 & 1.1 & 1.2 & 1.1 & {\bf 1.2} & 1.1 \\
 & V2 + V3 & 0.7 & 0.2 & 0.7 & 0.2 & 0.4 & 0.2 & {\bf 0.3} & 0.1 \\
\enddata
\end{deluxetable*}

\bibliography{BKxHIpaper}{}
\bibliographystyle{aasjournal}

\end{document}